\documentclass[lettersize,journal, twocolumn]{IEEEtran}
\usepackage{amsmath,amsfonts}
\usepackage{algorithmic}
\usepackage{algorithm}
\usepackage{array}
\usepackage{textcomp}
\usepackage{stfloats}
\usepackage{url}
\usepackage{verbatim}
\usepackage{graphicx}
\usepackage{cite}
\usepackage{tcolorbox}
\usepackage[hidelinks]{hyperref}
\usepackage{subcaption}
\usepackage{amsthm}
\usepackage{amssymb}
\theoremstyle{plain}
\newtheorem{theorem}{Theorem}[section]
\newtheorem{lemma}[theorem]{Lemma}
\newtheorem{proposition}[theorem]{Proposition}

\theoremstyle{definition}

\theoremstyle{remark}

\begin{document}

\title{Federated Learning for Terahertz Wireless Communication}

\newcommand{\orcidiconOtb}{\href{https://orcid.org/0009-0008-3903-2268}{\includegraphics[scale=0.1]{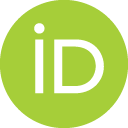}}}
\newcommand{\orcidiconOba}{\href{https://orcid.org/0000-0003-2523-3858}{\includegraphics[scale=0.1]{figures/orcidID128.png}}}

\author{O. Tansel Baydas\orcidiconOtb,~\IEEEmembership{Member,~IEEE}, Ozgur B. Akan\orcidiconOba,~\IEEEmembership{Fellow,~IEEE}
    \thanks{The authors are with  the Internet of Everything (IoE) Group, Electrical Engineering Division, Department of Engineering, University of Cambridge, Cambridge CB3 0FA, UK (e-mails: \{otb26, oba21\}@cam.ac.uk).}
    \thanks{O. B. Akan is also with the Center for neXt-generation Communications (CXC), Department of Electrical and Electronics Engineering, Ko\c{c} University, Istanbul 34450, Turkey (e-mail: akan@ku.edu.tr).}
}

\maketitle

\begin{abstract}
The convergence of Terahertz (THz) communications and Federated Learning (FL) promises ultra-fast distributed learning, yet the impact of realistic wideband impairments on optimization dynamics remains theoretically uncharacterized. This paper bridges this gap by developing a multicarrier stochastic framework that explicitly couples local gradient updates with frequency-selective THz effects, including beam squint, molecular absorption, and jitter. Our analysis uncovers a critical diversity trap: under standard unbiased aggregation, the convergence error floor is driven by the harmonic mean of subcarrier SNRs. Consequently, a single spectral hole caused by severe beam squint can render the entire bandwidth useless for reliable model updates. We further identify a fundamental bandwidth limit, revealing that expanding the spectrum beyond a critical point degrades convergence due to the integration of thermal noise and gain collapse at band edges. Finally, we demonstrate that an SNR-weighted aggregation strategy is necessary to suppress the variance singularity at these spectral holes, effectively recovering convergence in high-squint regimes where standard averaging fails. Numerical results validate the expected impact of the discussed physical layer parameters' on performance of THz-FL systems.

\end{abstract}


\section{Introduction}

\IEEEPARstart{N}{ext} generation wireless networks are envisioned to merge communication, computation, and intelligence into a unified ecosystem supporting extremely high data rates, low latency, and distributed intelligence. Among the candidate technologies enabling this vision, terahertz (THz) band communication and federated learning (FL) stand out as key pillars \cite{mcmahan2017communication}. The THz spectrum %(0.1--10~THz) 
offers unprecedented bandwidths capable of delivering multi-gigabit-per-second links, while FL enables distributed model training across multiple edge devices without exposing local data, thereby preserving privacy and reducing communication load \cite{akyildiz2014terahertz}.

THz band has been established as a core enabler of beyond-5G and 6G wireless systems \cite{ 7490372}. To meet connectivity, throughput, and latency requirements, future wireless systems will explore the wide bandwidth available at THz bands \cite{jiang2024terahertz}. \cite{jornet2011channel} introduced foundational channel models that incorporate unique physical-layer phenomena such as molecular absorption, spreading loss, and frequency-dependent path attenuation, setting the groundwork for understanding THz propagation characteristics.  
\cite{han2022terahertz, xue2024survey, sharma2025terahertz, 9766110} have consolidated the progress in device technologies, signal processing, and channel modeling, and have identified open challenges including beam misalignment, ultra-wideband beam-squint, and the need for high-dimensional array processing.  
Recent studies addressing resource allocation under THz impairments~\cite{7490372} also indicate that communication efficiency and coverage largely depend on distance-aware bandwidth and power control.

Parallel to the advances in THz communication, the concept of distributed learning over wireless networks has rapidly evolved. FL and its wireless application through over-the-air computation (AirComp) have emerged as efficient paradigms for edge intelligence \cite{amiri2020federated}. FL enables collaborative model training while transmitting only model parameters instead of raw data, drastically reducing communication overhead.  \cite{10799091} positions FL as a native component of next generation communication systems, emphasizing its dual role in the ``wireless for AI'' and ``AI for wireless'' paradigms. However, most existing studies focus on optimizing resource allocation, compression, or aggregation strategies under lower-frequency wireless models.

Recent studies have begun to apply FL for the THz communication problems. \cite{elbir2022federated,elbir2023federated} proposed federated multi-task learning frameworks for THz channel and direction-of-arrival (DoA) estimation, FL can drastically reduce communication overhead and improve beamspace support alignment.  
Similarly, \cite{li2023federated} integrated FL with intelligent reflecting surface (IRS)-assisted unmanned aerial vehicle (UAV) communications using over-the-air aggregation to jointly optimize UAV trajectory, power allocation, and IRS phase shifts. Furthermore, \cite{mahmood2024analysis} explicitly analyzed THz propagation characteristics, such as molecular absorption, to optimize the link design for FL in 6G systems. Recent works have also successfully applied FL to solve THz-specific physical layer challenges, such as beam-split misalignment \cite{elbir2023federated} and resource allocation \cite{10726611}. These works demonstrate that FL can effectively enhance the adaptability and data efficiency of THz-enabled networks.  However, they utilize FL as a tool to optimize the wireless link. To the best of our knowledge, no existing work has theoretically analyzed the impact of wideband THz impairments on the learning behaviour of the FL algorithm itself.

Existing works typically assume idealized fading channels with independent noise models, ignoring the distinct characteristics of THz propagation that can introduce correlated multiplicative and additive distortions into the learning process. Additionally, they primarily develop algorithmic or architectural solutions, focusing on estimation or system optimization. As a result, there is currently no theoretical model that characterizes how these physical-layer effects translate into convergence rates, bias, or variance inflation in federated learning algorithms. This paper addresses this fundamental gap by developing a comprehensive theoretical framework for federated learning over THz-band channels. Our main contributions are summarized as follows:
\begin{itemize}
  \item We establish a multicarrier stochastic model that couples local stochastic gradient descent (SGD) dynamics with frequency-selective THz impairments, capturing the deterministic gain degradation alongside molecular absorption, pointing jitter, and beam squint.
  \item We derive a non-convex convergence theorem that reveals a decoupling of impairments.
  \item We formulate design inequalities that link physical parameters to learning accuracy.
  \item We translate these abstract parameters into concrete THz array models, proposing strategies to mitigate spectral holes and identifying feasible operating regimes for learning-centric networks.
\end{itemize}

The remainder of the paper is organized as follows: Sec.~\ref{sec:system} introduces the system and modeling assumptions. Sec.~\ref{sec:thzmodel} develops the THz-band channel model. Sec.~\ref{sec:error} analyzes the received update statistics, constructing per-client bias, variance, and concentration bounds together with a bias-compensation estimator.
Sec.~\ref{sec:theory} derives the non-convex convergence theorem. Sec.~\ref{sec:design} formulates design inequalities linking physical-layer parameters to target learning accuracy and round budget. Sec.~\ref{sec:experiments} presents experimental validations, and Sec.~\ref{sec:conclusions} concludes the paper.

\begin{figure}
    \centering
    \includegraphics[width=0.7\linewidth]{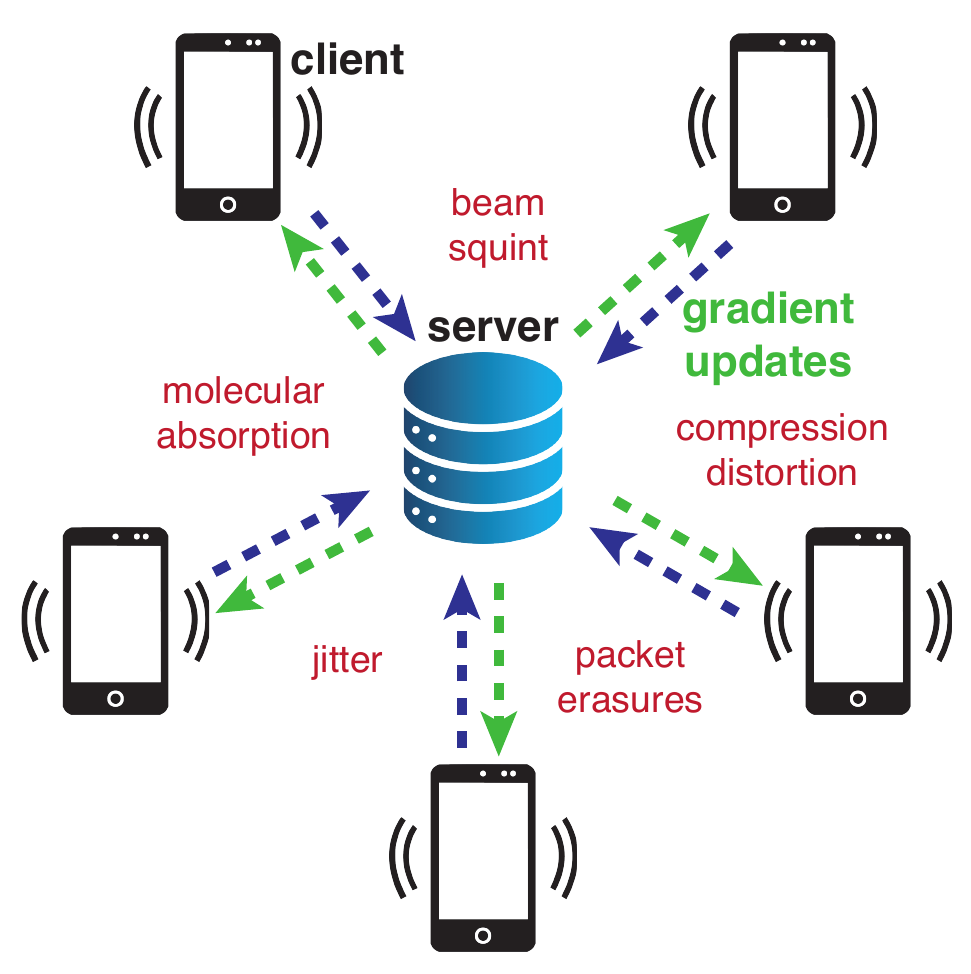}
    \caption{Topology and Challanges of FL in THz bands}
    \label{fig:placeholder}
\end{figure}

\section{System Model}
\label{sec:system}

\subsection{Federated Learning Topology}

We consider a standard synchronous federated learning (FL) system with a 
star topology. A central server maintains a global model $w_t\in\mathbb{R}^d$ 
at communication round $t$. There are $N$ clients indexed by 
$i\in\{1,\dots,N\}$, each holding a private data distribution $\mathcal D_i$ 
and corresponding empirical loss
\vspace{-0.5em}
\begin{equation}
    f_i(w_t) \;=\; \mathbb{E}_{\xi\sim\mathcal D_i}\!\left[\ell(w_t;\xi)\right].
\end{equation}
The global objective is the uniform average
\vspace{-0.5em}
\begin{equation}
    f(w_t)
    \;=\;
    \frac{1}{N}\sum_{i=1}^N f_i(w_t).
\end{equation}

At the start of round $t$, the server broadcasts $w_t$ to a subset of clients. We assume the downlink broadcast of the global model is error-free (e.g., via high-power beamforming at the BS) to focus on the uplink channel bottlenecks. Let $\mathcal S_t \subseteq \{1,\dots,N\}$ denote the selected set with $|\mathcal S_t| = m_t$. Each selected client performs $K$ steps of local SGD using a local stepsize $\eta_{\mathrm{loc}}$:
\vspace{-0.5em}
\[
w^{(i)}_{t,0}=w_t,
\qquad
w^{(i)}_{t,k+1}
=
w^{(i)}_{t,k}
-
\eta_{\mathrm{loc}}\,g^{(i)}_{t,k},
\]
where $g^{(i)}_{t,k}$ is a stochastic gradient of $f_i$ at $w^{(i)}_{t,k}$. 
After $K$ local steps, client $i$ forms its local update
\vspace{-0.5em}
\begin{equation}
    \Delta_{i,t}
    \;=\;
    w^{(i)}_{t,K} - w_t.
\end{equation}

The server receives distorted updates 
$\widetilde{\Delta}_{i,t}$ through THz uplinks and applies an aggregation rule
\vspace{-0.5em}

\begin{equation}
    \label{eq:unweighted}
    w_{t+1}
    \;=\;
    w_t
    \;+\;
    \eta_t\,
    \frac{1}{m_t}
    \sum_{i\in\mathcal S_t}
    \widetilde{\Delta}_{i,t},
    \vspace{-0.5em}
\end{equation}

\noindent where $\eta_t$ is the server stepsize. To support the convergence analysis, we impose the following standard assumptions 
on the losses and stochastic gradients: A1 (Smoothness), A2 (Bounded stochastic variance), A3 (Bounded heterogeneity):

\vspace{-1em}
\begin{align}
\text{A1:}\;&
\|\nabla f_i(x)-\nabla f_i(y)\| \le L\|x-y\|, \\[3pt]
\text{A2:}\;&
\mathbb{E}\|\nabla \ell(w_t;\xi)-\nabla f_i(w_t)\|^2 \le \sigma_i^2, \\[3pt]
\text{A3:}\;&
\frac1N \sum_{i=1}^N \|\nabla f_i(w_t)-\nabla f(w_t)\|^2 \le G^2.
\vspace{-1em}
\end{align}

% =====================================================
\subsection{Probability Space and Filtration}
\label{subsec:filtration}

All random quantities are defined on a probability space 
$(\Omega,\mathcal F,\mathbb P)$. For each $t\ge 0$, we let $\mathcal F_t$ 
denote the history of the system up to round $t$, including all randomness 
generated by local SGD, channel realizations, compression, and erasures:
\vspace{-0.5em}
\begin{equation}
\mathcal F_t
=
\sigma\!\left(
w_0,\;
\{\mathcal S_s,\;H_{i,s},\;d_{i,s},\;e_{c,i,s},\;\epsilon_{i,s}\}_{i\le N,\;s\le t}
\right),
\end{equation}
\noindent where SGD randomness up to $t$. By construction, $w_t$ and all local iterates are $\mathcal F_t$-measurable. The following assumptions describe the stochastic structure of the wireless-communication and local-computation processes.

\noindent \textit{a) Client sampling:}
The subset $\mathcal S_t$ is independent of $\mathcal F_{t-1}$ and is either i.i.d. drawn\ across rounds or uniformly sampled without replacement.

\noindent \textit{b) Block fading and multiplicative distortion:}
Given $\mathcal F_{t-1}$, the effective uplink channel gains 
$\{H_{i,t}\}_{i\le N}$ are independent across clients with
\vspace{-0.5em}
\[
\mathbb E[H_{i,t}] = \mu_{H,i},
\qquad
\mathrm{Var}\!\left(\frac{H_{i,t}}{\mu_{H,i}}\right)=\sigma_H^2.
\] 
The parameter $\sigma_H^2$ aggregates multiplicative distortions due to 
misalignment jitter, small-scale fading, and frequency-dependent beam-squint.

\noindent \textit{c) Packet erasures:}
Conditioned on $(H_{i,t}, \mathcal F_{t-1})$, each delivery indicator 
$d_{i,t}\in\{0,1\}$ is independent across clients with  
$\bar d_{i,t} = \mathbb E[d_{i,t}\mid H_{i,t}]$.

\noindent \textit{d)Compression noise:}
Given $(\Delta_{i,t}, \mathcal F_{t-1})$, the compression error $e_{c,i,t}$ is 
independent across clients, unbiased, and bounded in mean-square:
\[
\mathbb E[e_{c,i,t}\mid \Delta_{i,t}]=0,
\qquad
\mathbb E\|e_{c,i,t}\|^2
    \le \omega_t\,\|\Delta_{i,t}\|^2,
\]
where $\omega_t$ is determined by the quantizer bit budget or sparsification level.

\noindent \textit{e) Additive THz-induced distortion:}
Conditioned on $(H_{i,t},\mathcal F_{t-1})$, the additive noise 
$\epsilon_{i,t}$ is independent across clients and $\sigma_\epsilon$-sub-Gaussian \cite{goldsmith2005wireless}, 
satisfying:
\[
\mathbb E[\epsilon_{i,t}]\!=\!0,
\qquad 
\mathbb E[\exp(\lambda \epsilon_{i,t})] \!\le\! \exp\left(\!\frac{\lambda^2 \sigma_\epsilon^2}{2}\!\right) \quad \forall \!\lambda \!\in\! \mathbb{R}.
\]
This implies the variance bound $\mathbb E\|\epsilon_{i,t}\|^2 \le \sigma_\epsilon^2$ 
and captures residual symbol errors, thermal noise, and post-decoding distortion.

\noindent \textit{f) Local SGD noise independence:}
The stochastic gradients used in local updates at round $t$ are 
independent of $\{H_{i,t},d_{i,t},e_{c,i,t},\epsilon_{i,t}\}_i$ 
given $w_t$. This ensures that communication noise does not affect the 
local update directions.

\section{THz Channel Model and Communication Impairments}
\label{sec:thzmodel}

Each client communicates its local update $\Delta_{i,t}$ to the server over a 
wideband THz uplink. To combat frequency-selective fading and beam-squint, 
we assume the system utilizes Orthogonal Frequency Division Multiplexing (OFDM) \cite{goldsmith2005wireless}. 
The available bandwidth is divided into $N_c$ subcarriers. The model update 
$\Delta_{i,t}$ is partitioned into $N_c$ sub-vectors 
$\{\Delta_{i,t}^{(n)}\}_{n=1}^{N_c}$, where $\Delta_{i,t}^{(n)}$ is transmitted 
over the $n$-th subcarrier.

THz propagation introduces several distortions. This section formalizes each component under this multicarrier framework providing channel properties. 

\subsection{Deterministic Path Loss and Molecular Absorption}

For a client at distance $d_i$, the deterministic path gain at the 
frequency $f_n$ of the $n$-th subcarrier is \cite{jornet2011channel}
\vspace{-0.5em}
\begin{equation}
A_i(f_n,d_i)
=
\frac{c^2}{(4\pi f_n d_i)^2}
\exp\!\big(-k_a(f_n)\, d_i\big),
\label{eq:absorption}
\end{equation}
where $c$ denotes the speed of light and $k_a(f_n)$ is the 
frequency-dependent molecular absorption coefficient. This term captures 
the frequency-selective absorption characteristic of THz propagation, 
which varies across the $N_c$ subcarriers.

\subsection{Stochastic Gain Variations: Jitter, Fading, and Squint}
\label{subsec:gain-model}

At round $t$, the effective gain applied to the sub-vector 
$\Delta_{i,t}^{(n)}$ on the $n$-th subcarrier is modeled as \cite{8610080}:
\begin{equation}
H_{i,t}^{(n)}
=
g_i(f_n, d_i)\, h_{m,i,t}^{(n)}\, h_{f,i,t}^{(n)},
\end{equation}
where $g_i(f_n, d_i)$ incorporates the frequency-dependent path loss $A_i(f_n,d_i)$ and antenna gains, $h_{m,i,t}^{(n)}$ models random pointing-jitter and beam misalignment, and $h_{f,i,t}^{(n)}$ models small-scale fading on subcarrier $n$.

We define the mean gain for subcarrier $n$ as $\mu_{H,i}^{(n)}=\mathbb E[H_{i,t}^{(n)}]$ 
and the normalized fluctuation
\[
\widetilde H_{i,t}^{(n)}
=
\frac{H_{i,t}^{(n)}}{\mu_{H,i}^{(n)}},
\qquad
\mathbb E[\widetilde H_{i,t}^{(n)}]=1.
\]
Due to beam squint, $\mu_{H,i}^{(n)}$ may degrade significantly 
for subcarriers far from the center frequency. We aggregate the variance contributions into the effective multiplicative 
distortion parameter for subcarrier $n$:
\vspace{-0.5em}
\begin{equation}
\sigma_{H,n}^2
=
\mathrm{Var}(\widetilde H_{i,t}^{(n)}).
\label{eq:sigmaH-decomp}
\end{equation}

% ---------------------------------------------------------
\subsection{Compression Noise, Additive Distortion, and Erasures}

Compression is applied to the full vector, or independently per block. 
For analytical tractability, we model the compression error $e_{c,i,t}^{(n)}$ 
on the $n$-th sub-vector satisfying:
\vspace{-0.5em}
\begin{align}
\mathbb E[e_{c,i,t}^{(n)}\mid \Delta_{i,t}^{(n)}] &= 0, \\
\mathbb E\|e_{c,i,t}^{(n)}\|^2 &\le \omega_t\,\|\Delta_{i,t}^{(n)}\|^2.
\end{align}

Transmission introduces additive distortion $\epsilon_{i,t}^{(n)}$ on each 
subcarrier (thermal noise), which is bounded:
\[
\mathbb E[\epsilon_{i,t}^{(n)}] = 0,
\qquad 
\mathbb E\|\epsilon_{i,t}^{(n)}\|^2 \le \sigma_{\epsilon,n}^2.
\]

We model packet delivery using a Bernoulli indicator $d_{i,t}\in\{0,1\}$ 
assuming a block-fading erasure channel (all subcarriers succeed or fail together 
due to blockage): $\mathbb{P}(d_{i,t}=0)=p_{\mathrm{drop},i,t}$ and $\bar d_{i,t}=1-p_{\mathrm{drop},i,t}$.
% ---------------------------------------------------------

\subsection{Received Update Model}
\label{subsec:received-model}

To mitigate the impact of bursty spectral errors (structural erasure), we apply a random permutation matrix $\Pi_t$ to the update vector before subcarrier mapping. Let the permuted update be $\Delta_{i,t}^{\pi} = \Pi_t \Delta_{i,t}$. This vector is partitioned into sub-vectors $\{\Delta_{i,t}^{\pi,(n)}\}$. Putting all terms together, the server receives the $n$-th sub-vector as:
\begin{equation}
\widetilde{\Delta}_{i,t}^{(n)}
=
d_{i,t}\Big(H_{i,t}^{(n)}(\Delta_{i,t}^{\pi,(n)}+e_{c,i,t}^{(n)}) + \epsilon_{i,t}^{(n)}\Big).
\label{eq:received}
\end{equation}
Factoring out the subcarrier-specific mean $\mu_{H,i}^{(n)}$ reveals the normalized fluctuation structure:
\begin{equation}
\begin{split}
\widetilde{\Delta}_{i,t}^{(n)}
&= d_{i,t}\mu_{H,i}^{(n)}\!\left(
\Delta_{i,t}^{\pi,(n)} + e_{c,i,t}^{(n)}
\right. \\
&\quad {}+ (\widetilde H_{i,t}^{(n)}-1)(\Delta_{i,t}^{\pi,(n)}+e_{c,i,t}^{(n)})
\left.
+ \epsilon'^{(n)}_{i,t}
\right),
\end{split}
\label{eq:normalized-received}
\end{equation}

with $\epsilon'^{(n)}_{i,t}=\epsilon_{i,t}^{(n)}/\mu_{H,i}^{(n)}$.

\subsection{Mean-Gain Estimation and Compensation}
\label{subsec:compensation}

To correct for bias, the server estimates the effective mean gain matrix. 
Assuming pilot signals are available on all subcarriers, the server computes 
$\widehat\theta_{i,t}^{(n)}$ for each $n$, tracking $\theta_i^{\star(n)} \!=\! \bar d_i \mu_{H,i}^{(n)}$. The compensated update is formed by element-wise rescaling:
\vspace{-0.5em}
\[
\widetilde{\Delta}^{(\mathrm{scaled})}_{i,t}
=
\bigoplus_{n=1}^{N_c} \left( \frac{1}{\widehat\theta_{i,t}^{(n)}} \,\widetilde{\Delta}_{i,t}^{(n)} \right).
\] \vspace{-0.5em}

By Lemma~\ref{lem:theta-concentration}, assuming sufficient pilot density $M$, 
the dominant mean bias is removed per-subcarrier.

\begin{lemma}[Concentration of the gain estimator]
\label{lem:theta-concentration}
Let the estimator for subcarrier $n$ be $\widehat{\theta}_{i,t}^{(n)}=\frac{1}{M}\sum_{s=t-M+1}^{t} z_{i,s}^{(n)}$, 
where $z_{i,s}^{(n)}$ are unbiased instantaneous pilot statistics. Assume the pilot noise is conditionally sub-Gaussian 
with proxy variance $\nu_i^2$. Then for any $\epsilon>0$:
\vspace{-0.5em}
\[
\mathbb P\left(\left|\widehat\theta_{i,t}^{(n)} - \bar d_i \mu_{H,i}^{(n)}\right| > \epsilon\right)
\;\le\;
2\exp\left(-\frac{M\epsilon^2}{2\nu_i^2}\right).
\]
In particular, $\widehat\theta_{i,t}^{(n)} \xrightarrow{p} \bar d_i\mu_{H,i}^{(n)}$ as $M\to\infty$. \begin{footnote}{
While Lemma~\ref{lem:theta-concentration} guarantees convergence, the validity of the channel compensation scheme in Section~\ref{sec:thzmodel} requires that the true mean gain is bounded away from zero, i.e., $\bar d_i \mu_{H,i}^{(n)} \ge \delta > 0$. Subcarriers violating this condition (e.g., deep beam squint nulls) must be excluded from the aggregation (packet loss) to prevent the noise amplification term $\frac{\sigma_{\epsilon,n}^2}{(\mu_{H,i}^{(n)})^2}$ from diverging.}
\end{footnote}
\end{lemma}

\subsection{Validity of the OFDM Model}
\label{subsec:scalar-validity}

By employing OFDM, the wideband frequency-selective channel is 
decomposed into $N_c$ parallel narrowband subchannels. The scalar gain 
approximation $H_{i,t}^{(n)}$ is valid within each subcarrier bandwidth $\Delta f$ 
provided $\Delta f \ll f_c$. This resolves the beam-squint limitation of 
single-carrier models, as gain variations across the bandwidth are explicitly 
captured by the variation of $\mu_{H,i}^{(n)}$ across $n$.

\subsection{Statistical Abstraction of the Channel}
\label{subsec:comm-assumptions}

To facilitate the convergence analysis in the following sections, we distill the physical THz model into the following statistical properties:

\noindent \textit{a) Per-subcarrier Multiplicative Noise:}
For each subcarrier $n$, the channel gain is modeled as $H_{i,t}^{(n)} = \mu_{H,i}^{(n)}(1+\widetilde H_{i,t}^{(n)})$, where $\mu_{H,i}^{(n)}$ is the deterministic mean gain capturing path loss and beam squint, and $\widetilde H_{i,t}^{(n)}$ is a zero-mean fluctuation with variance $\sigma_{H,n}^2$.

\noindent \textit{b) Independence Structure:}
Channel realizations are independent across clients. Across subcarriers, fading may be correlated; however, our worst-case stability analysis relies on the maximum jitter variance $\max_n \sigma_{H,n}^2$, which holds valid regardless of spectral correlation.

\noindent \textit{c) Frequency-Selective Additive Noise:}
The additive distortion is bounded per subcarrier: $\mathbb E\|\epsilon_{i,t}^{(n)}\|^2 \le \sigma_{\epsilon,n}^2$. 
Crucially, this variance is not uniform; it scales inversely with the allocated power $P_i^{(n)}$ and determines the error floor via the harmonic mean of the resulting SNRs.

\noindent \textit{d) Whitening via Parameter Interleaving:} Wideband THz channels introduce structured distortions, including deterministic beam squint and correlated pointing jitter. To prevent these from translating into correlated gradient noise, clients employ parameter interleaving (random permutation) \cite{goldsmith2005wireless}. This acts as statistical whitening, converting localized spectral correlations into independent, isotropic noise across coordinates. Consequently, our convergence theorems can validly treat THz impairments as independent per-coordinate perturbations.

\section{Error Decomposition and Statistical Structure}
\label{sec:error}

We now isolate the contribution of the THz link. The total communication error is 
the concatenation of errors on all subcarriers:
\vspace{-0.5em}
\begin{equation}
E^{\mathrm{comm}}_{i,t}
\;\triangleq\;
\widetilde{\Delta}_{i,t} - \Delta_{i,t} 
= \bigoplus_{n=1}^{N_c} (\widetilde{\Delta}_{i,t}^{(n)} - \Delta_{i,t}^{(n)}).
\end{equation}

% ---------------------------------------------------------
\subsection{Bias}

Using the per-subcarrier compensation $\widehat\theta_{i,t}^{(n)}$, the server effectively inverts the channel gain. Consequently, the expected compensated update becomes unbiased with respect to the channel realization, subject only to packet erasures:
\vspace{-0.5em}
\begin{equation}
\mathbb E[\widetilde{\Delta}_{i,t}^{(n)} \mid \mathcal F_{t-1}]
=
\bar d_{i,t}\,
\mathbb E[\Delta_{i,t}^{(n)} \mid \mathcal F_{t-1}].
\label{eq:bias-corrected}
\end{equation}
The channel gain $\mu_{H,i}^{(n)}$ does not appear in eq.\eqref{eq:bias-corrected} because it has been removed by the estimator $\widehat\theta_{i,t}^{(n)} \approx \bar{d}_i \mu_{H,i}^{(n)}$. The remaining bias $-(1-\bar d_{i,t})\Delta_{i,t}^{(n)}$ arises from the probability of packet loss.

% ---------------------------------------------------------
\subsection{Frequency-Selective Second-order Error}

We expand the squared norm of the error. In the OFDM regime, 
we sum the errors over the subcarriers:
\vspace{-0.5em}
\begin{equation}
\mathbb E\|E^{\mathrm{comm}}_{i,t}\|^2
=
\sum_{n=1}^{N_c} \mathbb E \|\widetilde{\Delta}_{i,t}^{(n)} - \Delta_{i,t}^{(n)}\|^2.
\end{equation}

\noindent Substituting the compensated model from eq.~\eqref{eq:normalized-received} into the sum:
\vspace{-0.6em}
\begin{align}
\mathbb E\|E^{\mathrm{comm}}_{i,t}\|^2
&\le
\sum_{n=1}^{N_c} 
\bigg[
\underbrace{\big(1-\bar d_{i,t}\big)^2
\,\mathbb E\|\Delta_{i,t}^{(n)}\|^2}_{\text{Residual Bias (Erasures)}}
\notag\\[-2pt]
&\quad+
\bar d_{i,t}
\bigg(
(\sigma_{H,n}^2+\omega)\,\mathbb E\|\Delta_{i,t}^{(n)}\|^2
+
\frac{\sigma_{\epsilon,n}^2}{(\mu_{H,i}^{(n)})^2}
\bigg)
\bigg],
\label{eq:comm-variance-tight}
\end{align}

\noindent where the first term captures the bias due to packet erasures, while the second term captures the compensated variance contributions of channel jitter/fading ($\sigma_{H,n}^2$), compression ($\omega$), and additive noise ($\sigma_{\epsilon,n}^2$).

Eq.~\eqref{eq:comm-variance-tight} reveals the critical impact of beam squint and frequency selectivity. The error here depends on the \emph{alignment} between the model importance and channel quality. 
Specifically, if a sub-vector $\Delta_{i,t}^{(n)}$ is mapped to a subcarrier $n$ with low gain (small $\mu_{H,i}^{(n)}$), the additive noise variance is amplified by the factor $1/(\mu_{H,i}^{(n)})^2$. Similarly, mapping gradients to subcarriers with high instability (large $\sigma_{H,n}^2$) inflates the multiplicative error. This motivates the use of interleaving or channel-aware scheduling to avoid placing critical parameters on deep-faded or unstable frequencies.

\section{Aggregate Statistics}
\label{sec:theory}

For a participating client set $\mathcal S_t$ of size $m_t$, the server forms 
the %pre-stepsize 
aggregated update. Under the multicarrier model, this aggregation 
occurs per subcarrier:
\vspace{-0.5em}
\begin{equation}
\widehat{\Delta}_t^{(n)}
=
\frac{1}{m_t}
\sum_{i\in\mathcal S_t}
\big(\Delta_{i,t}^{(n)} + E^{\mathrm{comm}}_{i,t,n}\big),
\label{eq:agg-def}
\end{equation}
where $E^{\mathrm{comm}}_{i,t,n} = \widetilde{\Delta}_{i,t}^{(n)} - \Delta_{i,t}^{(n)}$ 
is the error on the $n$-th subcarrier. The full aggregated update is the 
concatenation $\widehat{\Delta}_t = \bigoplus_{n} \widehat{\Delta}_t^{(n)}$.

% ---------------------------------------------------------
\subsection{Mean and Bias}

With the per-subcarrier compensation 
described in Section~\ref{sec:error}, the estimator for each sub-vector is 
asymptotically unbiased:
\vspace{-0.5em}
\[
\mathbb E[\widehat{\Delta}_t^{(n)}\mid \mathcal F_{t-1}]
=
\overline{g}_t^{(n)} + O(\epsilon),
\]
where $\overline{g}_t^{(n)}$ is the average of the true local sub-vectors 
$\Delta_{i,t}^{(n)}$ and $O(\epsilon)$ is the negligible estimation error. Consequently, the full vector $\widehat{\Delta}_t$ is 
unbiased up to the concatenation of these small residual errors.

% ---------------------------------------------------------
\subsection{Variance}

The total variance of the aggregated update is the sum of the variances across 
all subcarriers. Using conditional independence across clients and subcarriers, we have: 
\vspace{-0.5em}
\begin{equation}
\mathbb E\|\widehat{\Delta}_t \!-\! \mathbb E[\widehat{\Delta}_t\mid\mathcal F_{t-1}]\|^2
\!\;\le\;\!
\frac{1}{m_t}
\Big(
V_{\mathrm{loc},t}
\!+\!
V_{\mathrm{het},t}
\!+\!
V_{\mathrm{comm},t}
\Big).
\label{eq:agg-var}
\end{equation}

\noindent where $V_{\mathrm{loc},t}, V_{\mathrm{het},t}$, and $V_{\mathrm{comm},t}$ are local SGD, data heterogeneity (non-i.i.d.), and communication variances, respectively. $V_{\mathrm{comm},t}$ now aggregates distortions 
across the frequency band. Summing the bound over $i \in \mathcal S_t$ yields:
\vspace{-0.5em}
\begin{align}
V_{\mathrm{comm},t}\!
\lesssim\!
\frac{1}{m_t}\!
\sum_{i\in\mathcal S_t}\!
\sum_{n=1}^{N_c} \!
\bar d_{i,t} \!
\left[
(\sigma_{H,n}^2\!+\!\omega)\,\mathbb E\|\Delta_{i,t}^{(n)}\|^2
+ \frac{\sigma_{\epsilon,n}^2}{(\mu_{H,i}^{(n)})^2}
\right]\!.
\label{eq:comm-variance}
\end{align}

\begin{lemma}[Explicit bound on the local update energy]
\label{lem:local-energy-explicit}
Assume each $f_i$ is $L$-smooth and local SGD uses stepsize $\eta_{\text{loc}}\le \frac{1}{2L}$ with mini-batch noise variance bounded by $\sigma_{\text{sgd}}^2$. 
Let $G$ be the bound on gradient heterogeneity, i.e., $\|\nabla f_i(w) - \nabla f(w)\| \le G$.
Then there exist absolute constants $a_1, a_2, a_3$ 
such that for every participating client $i$ at round $t$:
\begin{equation}
\mathbb E\|\Delta_{i,t}\|^2
\;\le\;
8\,\eta_{\text{loc}}^2 K^2\,\|\nabla f(w_t)\|^2
+ 2\,\eta_{\text{loc}}^2 K\,\sigma_{\text{sgd}}^2
+ 8\,\eta_{\text{loc}}^2 K^2\,G^2.
\end{equation}
This applies to the norm of the full vector $\|\Delta_{i,t}\|^2 = \sum_n \|\Delta_{i,t}^{(n)}\|^2$.
\end{lemma}

This explicit summation highlights the risk of frequency-selective fading: 
if a specific subcarrier $n$ has high jitter ($\sigma_{H,n}^2$), it distorts the signal multiplicatively. 
Also, subcarriers with low gain $\mu_{H,i}^{(n)}$ due to beam squint cause noise amplification during channel compensation, causing the effective noise term $\sigma_{\epsilon,n}^2/(\mu_{H,i}^{(n)})^2$ to explode for the corresponding model parameters $\Delta_{i,t}^{(n)}$.

\subsection{Non-convex Convergence}

The following theorem provides the fundamental stability analysis for the proposed multicarrier framework. It proves that despite the severe fluctuations caused by beam squint and pointing errors, the algorithm achieves an $\mathcal{O}(1/\sqrt{T})$ convergence rate, provided the physical layer parameters satisfy specific variance constraints.

\begin{theorem}[Non-convex convergence under Multicarrier THz impairments]
\label{thm:main}
Assume mean-bias compensation, parameter interleaving, constant stepsizes $\eta_t=\eta$ and 
$m_t=m$, and small local stepsizes such that $\eta_{\mathrm{loc}}K\le c_0/L$.  
Then:
\vspace{-0.5em}
\begin{align}
\frac{1}{T}\sum_{t=0}^{T-1}
\mathbb E\|\nabla f(w_t)\|^2
&\le
\frac{2\big(f(w_0)-f^\star\big)}
{\eta\,\eta_{\mathrm{loc}}c_0KT}
+
\frac{c_1\eta L}{m}\sigma_{\mathrm{sgd}}^2
\notag\\[-3pt]
&\quad+
c_2\eta L\,\Phi_{\mathrm{THz}}^{\mathrm{MC}}
+
c_3\eta^2L^2K\,G^2,
\label{eq:nonconvex-rate}
\end{align}
where the Multicarrier THz penalty $\Phi_{\mathrm{THz}}^{\mathrm{MC}}$ is defined as:
\vspace{-0.5em}
\begin{align}
\Phi_{\mathrm{THz}}^{\mathrm{MC}}
&=
\frac{1}{m} \Bigg[
\sum_{n=1}^{N_c} \frac{\sigma_{\epsilon,n}^2}{(\bar\mu_H^{(n)})^2}
\notag\\[-2pt]
&\quad+
\left(\! \frac{1}{N_c}\!\sum_{n=1}^{N_c} (\sigma_{H,n}^2\!+\!\omega)\! \right)
\!\big(\!a_2\eta_{\mathrm{loc}}^2K\sigma_{\mathrm{sgd}}^2
\!+\!a_3\eta_{\mathrm{loc}}^2K^2G^2\!\big)
\!\Bigg].
\label{eq:phi-thz}
\end{align}
Here $\bar\mu_{H}^{(n)}$ is the average gain of subcarrier $n$ across clients. With parameter interleaving, the multiplicative penalty scales with the \emph{arithmetic mean} of the distortion variances across the band, significantly relaxing the stability constraints compared to sequential mapping.\footnote{Parameter interleaving reduces the worst-case dependence only for multiplicative noise, because permutation averaging eliminates spectral correlation. In contrast, the additive distortion term depends on the inverse channel gains; beam squint on specific subcarriers can still dominate the sum of inverse gains.}
\end{theorem}

The bound in \eqref{eq:nonconvex-rate} confirms that with $\eta \propto 1/\sqrt{T}$, the algorithm reaches a stationary point at a rate of $\mathcal{O}(1/\sqrt{T})$. However, the convergence floor is dominated by the harmonic mean of the subcarrier SNRs due to the implicit assumption of unbiased channel inversion. As $\mu_{H,i}^{(n)} \to 0$ (e.g., severe beam squint), the noise amplification term $\sigma_{\epsilon,n}^2/(\mu_{H,i}^{(n)})^2$ diverges. To mitigate this ``Spectral Hole'' bottleneck, we propose an SNR-weighted aggregation scheme. Instead of uniform averaging, the server aggregates subcarrier $n$ using weights $\alpha_{i,t}^{(n)}$ inversely proportional to the effective noise variance:
\vspace{-0.5em}
\begin{equation}
    \hat{\Delta}_t^{(n)} = \frac{1}{\sum_{j \in \mathcal{S}_t} \alpha_{j,t}^{(n)}} \sum_{i \in \mathcal{S}_t} \alpha_{i,t}^{(n)} \left( \frac{\tilde{\Delta}_{i,t}^{(n)}}{\hat{\theta}_{i,t}^{(n)}} \right), 
    \label{eq:weighted}
\end{equation}
where $\alpha_{i,t}^{(n)}$ is defined as $\!\propto\! \left(\! \sigma_{\epsilon,n}^2/(\mu_{H,i}^{(n)})^2 \!+\! \sigma_{H,n}^2\!+\! \omega \!\right)^{-1}$. This down-weights clients experiencing deep fading or high jitter, preventing the variance term in $\Phi_{\mathrm{THz}}^{\mathrm{MC}}$ from diverging. While this stabilizes learning, it introduces an implicit objective bias. The scheme effectively optimizes a surrogate objective $f_{\boldsymbol{\alpha}}(w) = \sum \bar{\alpha}_i f_i(w)$, where $\bar{\alpha}_i$ are the normalized weights. The convergence behavior follows a Bias-Variance Tradeoff:
\vspace{-0.5em}
\begin{equation}
\min_{t < T} \mathbb{E}\|\nabla f(w_t)\|^2
\!\;\le\;\!
\underbrace{\mathcal{O}\!\left(\!\frac{1}{\sqrt{T}}\!\right)}_{\text{Decaying Variance}}
\!+\!
\underbrace{\mathcal{O}\left(\! \sum_{i=1}^N \left| \bar{\alpha}_i \!-\! \frac{1}{N} \right|^2 \!G^2 \!\right)}_{\text{Asymptotic Bias Floor}}.
\label{eq:bias_tradeoff}
\end{equation}
If we insist on unbiased aggregation ($\bar{\alpha}_i = 1/N$), the bias vanishes, but the variance term explodes due to the harmonic mean bottleneck. By accepting the bounded bias floor in \eqref{eq:bias_tradeoff}, we ensure the variance remains finite, recovering practical convergence in high-squint regimes as demonstrated in Sec.~\ref{sec:design}.

\section{Design Inequalities and Physical Parameterization}
\label{sec:design}

We now translate the convergence analysis into explicit engineering constraints for the multicarrier parameters. We fix a round budget $T$, local steps $K$, and stepsizes $\eta = \alpha/\sqrt{T}$, $\eta_{\mathrm{loc}} = \beta/\sqrt{K}$ (fixed local stepsize) such that the inflation condition holds. Let $\varepsilon > 0$ denote the target average stationarity. Substituting this schedule into Theorem~\ref{thm:main} yields the master bound:
\vspace{-0.5em}
\begin{align}
\frac{1}{T}\sum_{t=0}^{T-1}\mathbb E\|\nabla f(w_t)\|^2
\;\le\;&
\frac{C_{\mathrm{opt}}}{\sqrt{T}}
+ \frac{1}{\sqrt{T}}\left(C_{\mathrm{sgd}}\sigma_{\mathrm{sgd}}^2
+ C_{\mathrm{thz}}\Phi_{\mathrm{THz}}^{\mathrm{MC}}\right) \notag\\
&+\,\frac{C_{\mathrm{het}}G^2}{T},
\label{eq:design-bound-avg}
\end{align}
where $C_{\mathrm{opt}} = \frac{2(f(w_0)-f^\star)}{\alpha\beta c_0 \sqrt{K}}$, $C_{\mathrm{sgd}} = \frac{c_1 \alpha L}{m}$, $C_{\mathrm{thz}} = c_2 \alpha L$, and $C_{\mathrm{het}} = c_3 \alpha^2 L^2 K$. Note that the heterogeneity error decays faster ($1/T$) than the optimization and noise terms ($1/\sqrt{T}$).

\vspace{-0.5em}
\subsection{Multicarrier Design Inequalities}

The following proposition establishes the sufficient conditions for the system to achieve accuracy $\varepsilon$.

\begin{proposition}[Multicarrier Stability Conditions]
\label{prop:design-nonconvex}
To achieve the target accuracy $\varepsilon$, the specific error sources within $\Phi_{\mathrm{THz}}^{\mathrm{MC}}$ must satisfy the following budget constraints:

\noindent 1. Total SNR (Additive Noise): The integrated additive noise across the band is governed by the harmonic mean of the channel gains. This must satisfy:
\vspace{-0.3em}
\begin{align}
\frac{1}{m}\sum_{n=1}^{N_c} \frac{\sigma_{\epsilon,n}^2}{(\bar\mu_H^{(n)})^2}
\;\le\;
\frac{\varepsilon}{4C_{\mathrm{thz}}}\sqrt{T}.
\label{eq:design-snr}
\end{align}

\noindent 2. Average Stability: Due to interleaving, the average multiplicative distortion must not amplify local noise beyond the budget. Note the inclusion of the aggregation gain $1/m$:
\vspace{-0.3em}
\begin{multline}
\label{eq:design-jitter}
\frac{1}{m}
\left( \frac{1}{N_c}\sum_{n=1}^{N_c} (\sigma_{H,n}^2+\omega_n) \right)
\left(a_2\eta_{\mathrm{loc}}^2K\sigma_{\mathrm{sgd}}^2 + a_3\eta_{\mathrm{loc}}^2K^2G^2\right)
\\[-3pt]
\le\;
\frac{\varepsilon}{8C_{\mathrm{thz}}}\sqrt{T}.
\end{multline}

\noindent 3. Absorbability: To prevent divergence, the average jitter and compression must satisfy the descent-absorption condition:
\vspace{-0.3em}
\begin{align}
\frac{1}{m}\,
\left( \frac{1}{N_c}\sum_{n=1}^{N_c} (\sigma_{H,n}^2+\omega_n) \right)\;
a_1\eta_{\mathrm{loc}}^2K^2
\;\le\;
\frac{1}{4L}.
\label{eq:design-absorb}
\end{align}

\noindent 4. Standard FL Conditions: The round budget $T$ must suppress standard errors:
\vspace{-0.3em}
\begin{align}
\frac{C_{\mathrm{opt}}}{\sqrt{T}} \le \frac{\varepsilon}{4}, \quad
\frac{C_{\mathrm{sgd}}\sigma_{\mathrm{sgd}}^2}{\sqrt{T}} \le \frac{\varepsilon}{4}, \quad
\frac{C_{\mathrm{het}}G^2}{T} \le \frac{\varepsilon}{4}.
\end{align}
\end{proposition}

\subsection{Physical Instantiation and Scaling Insights}
\label{subsec:physical_instantiation}

We now instantiate the abstract terms in Proposition~\ref{prop:design-nonconvex} using physical array parameters to reveal the dependencies on bandwidth $B$, beam squint, and quantization.

\subsubsection{Beam Squint and Bandwidth Limit}
The additive noise condition eq.~\eqref{eq:design-snr} is bottlenecked by subcarriers with low gain. In wideband THz communication, the array gain is frequency-dependent due to beam squint. Physically, beam squint causes the spatial direction of the main lobe to shift as a function of frequency. For a subcarrier frequency $f_n$ and a target user angle $\phi_i$, the array gain is given by \cite{wang2019beam}:
\begin{equation}
\mu_{H,i}^{(n)} \propto \left| \operatorname{sinc}\left( \frac{N_{\text{ant}} \pi f_n d_{\text{ant}}}{c} (\sin \phi_i - \sin \phi_{\text{squint}}^{(n)}) \right) \right|^2,
\label{eq:squint-sinc}
\end{equation}

\noindent where, $\phi_{\text{squint}}^{(n)} = \arcsin(\frac{f_c}{f_n}\sin\phi_0)$ is the frequency-dependent pointing angle, $N_{\text{ant}}$ is number of antennas in the Uniform Linear Array (ULA), and $d_{\text{ant}}$ is antenna element spacing. While beam squint physically redirects signal energy (potentially towards other spatial locations), from the perspective of the intended client $i$, this manifests as a gain collapse on subcarriers where the beam misaligns with $\phi_i$. In a multi-user FL system, we model this as a worst-case per link degradation: even if the squinted beam accidentally aligns with another user, it does not carry the correct gradient information for that user, effectively rendering the subcarrier useless for reliable model aggregation.

As bandwidth $B$ expands, edge subcarriers deviate significantly from the center frequency, causing the main lobe to deviate away from the user angle $\phi_i$. Substituting the noise variance $\sigma_{\epsilon,n}^2 \approx \frac{d_{\mathrm{sub}}}{\kappa} (\mathrm{SNR}_{\mathrm{eff}}^{(n)})^{-1}$ into eq.~\eqref{eq:design-snr}, the constraint becomes:
\begin{equation}
\sum_{n=1}^{N_c} \frac{1}{(\bar\mu_{H}^{(n)})^2 \mathrm{SNR}_{\mathrm{eff}}^{(n)}}
\;\lesssim\;
\frac{m\,\varepsilon}{4 C_{\mathrm{thz}}} \frac{\kappa}{d_{\mathrm{sub}}} \sqrt{T}.
\label{eq:design-snr-closed}
\end{equation}

This reveals a fundamental physical limit: as edge subcarriers fall into squint nulls ($\bar\mu_{H}^{(n)} \to 0$), the LHS of eq.~\eqref{eq:design-snr-closed} diverges. Consequently, expanding the spectrum beyond a critical point degrades convergence. To maximize stability, transmit power should be allocated to equalize the weighted inverse-SNR term across the band.

\subsubsection{Jitter Sensitivity and Adaptive Quantization}
The absorbability condition eq.~\eqref{eq:design-absorb} limits the average multiplicative noise. The subcarrier variance $\sigma_{H,n}^2$ is not uniform; it depends on the jitter sensitivity $S_{\mathrm{sens}}$ (the slope of the array factor):
\begin{equation}
  \sigma_{H,n}^2
  \;\approx\;
   S_{\mathrm{sens}}(f_n) \, \sigma_{\theta}^2
  + \sigma_{\mathrm{fading}}^2.
  \label{eq:sigmaH_param}
\end{equation}
Edge subcarriers operate on the steep slope of the main lobe, resulting in high sensitivity $S_{\mathrm{sens}}$ to angular pointing errors $\sigma_\theta$.
We utilize the Quantized SGD (QSGD) relationship $\omega_n(b_n) \approx \min\{\frac{d_{\text{sub}}}{2^{2b_n}}, 1\}$, where $d_{\text{sub}}$ and $b_n$ are dimension of the sub-vector and quantization bit budget, respectively \cite{alistarh2017qsgd}. The stability constraint forces a tradeoff between physical jitter and digital compression:
\begin{equation}
\frac{1}{N_c}\sum_{n=1}^{N_c} \frac{d_{\text{sub}}}{2^{2b_n}}
\;\le\;
\frac{m}{4L a_1 \eta_{\mathrm{loc}}^2 K^2} - \bar{\sigma}_{H}^2.
\end{equation}

This inequality dictates the total bit budget. Since edge subcarriers exhibit high physical jitter $\sigma_{H,n}^2$, they consume a larger portion of the allowable variance budget. To satisfy the inequality, one must either increase the global bit budget or employ non-uniform quantization, allocating higher precision (larger $b_n$) to sensitive subcarriers to suppress their quantization noise $\omega_n$.

\section{Experimental Analyses}
\label{sec:experiments}

This section outlines the numerical studies required to empirically support 
the theoretical multicarrier framework of Sections~\ref{sec:theory}--\ref{sec:design}. 
The experiments are designed to: (1) validate the convergence behaviors under frequency-selective fading; (2) isolate the impact of wideband impairments (squint, jitter, thermal noise); (3) verify the necessity of channel compensation and equalization; (4) demonstrate the robustness of the proposed weighted aggregation strategy compared to standard averaging under severe channel stress.

\textit{Simulation Setup and Training Parameters:} We simulate a THz-FL environment consisting of $N\!=\!10$ clients and one central server. The underlying machine learning task is image classification using the MNIST dataset. The global model is a Convolutional Neural Network (CNN) comprising two convolutional layers and two fully connected layers. We utilize SGD with a learning rate $\eta = 0.02$, a batch size of 32, and a local epoch count of $E=1$. The total number of communication rounds is set to $T=10$. To simulate realistic wireless constraints, we enforce a spectral efficiency safety threshold of $\tau = 0.01$ bits/Hz; packets falling below this SNR floor are treated as erasures.

\textit{Physical Layer Parameters:} The carrier frequency is set to $f_c = 300$ GHz with a total bandwidth $B = 10$ GHz, divided into $N_c = 128$ subcarriers. The base station utilizes a ULA with $N_{ant}=64$ antennas. The channel model incorporates path loss, molecular absorption, and frequency-dependent array gains.

\begin{figure}[htp!]
    \centering
    \includegraphics[width=0.7\linewidth]{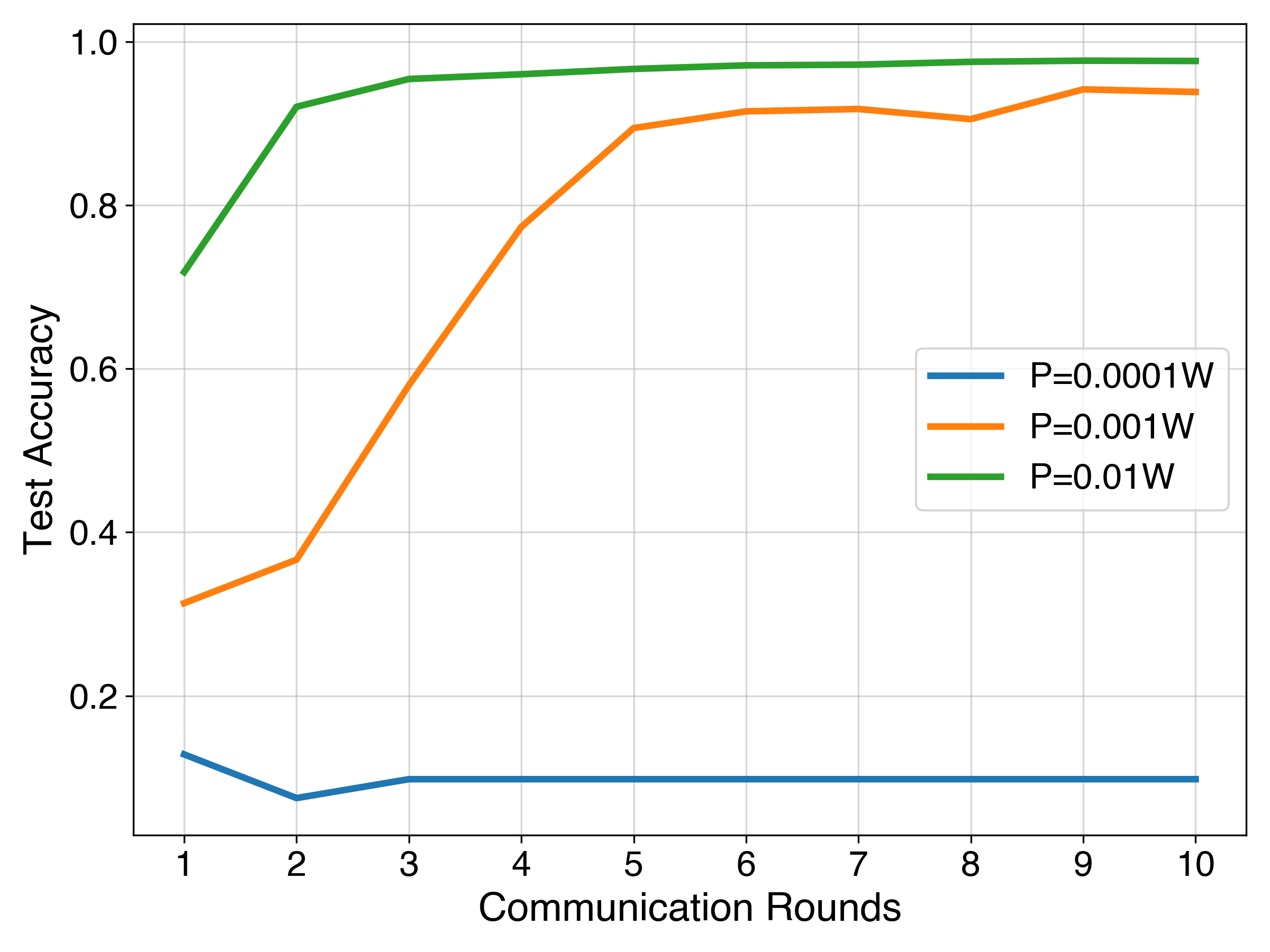}
    \caption{Impact of transmit power on convergence. Low power leads to signal erasure.}
    \label{fig:exp_a}
\end{figure}
\vspace{-0.5em}

\subsection{Impact of Transmit Power}
To establish the baseline sensitivity of the THz link, we evaluate the test accuracy under varying transmit powers $P \in \{0.1 \text{mW}, 10 \text{mW}, 1 \text{W}\}$ where power budget constrained in each test. As shown in Fig.~\ref{fig:exp_a}, a transmit power of $0.1$ mW fails to overcome the noise floor, resulting in continuous packet erasure and a stagnated accuracy of $\approx 9.8\%$ (random guessing). Conversely, powers of $10$ mW and above provide sufficient SNR to sustain a stable link, achieving $\approx 98\%$ accuracy.

\begin{figure*}[htp!]
    \centering
    
    \begin{subfigure}[b]{0.32\textwidth}
        \centering
        \includegraphics[width=\linewidth]{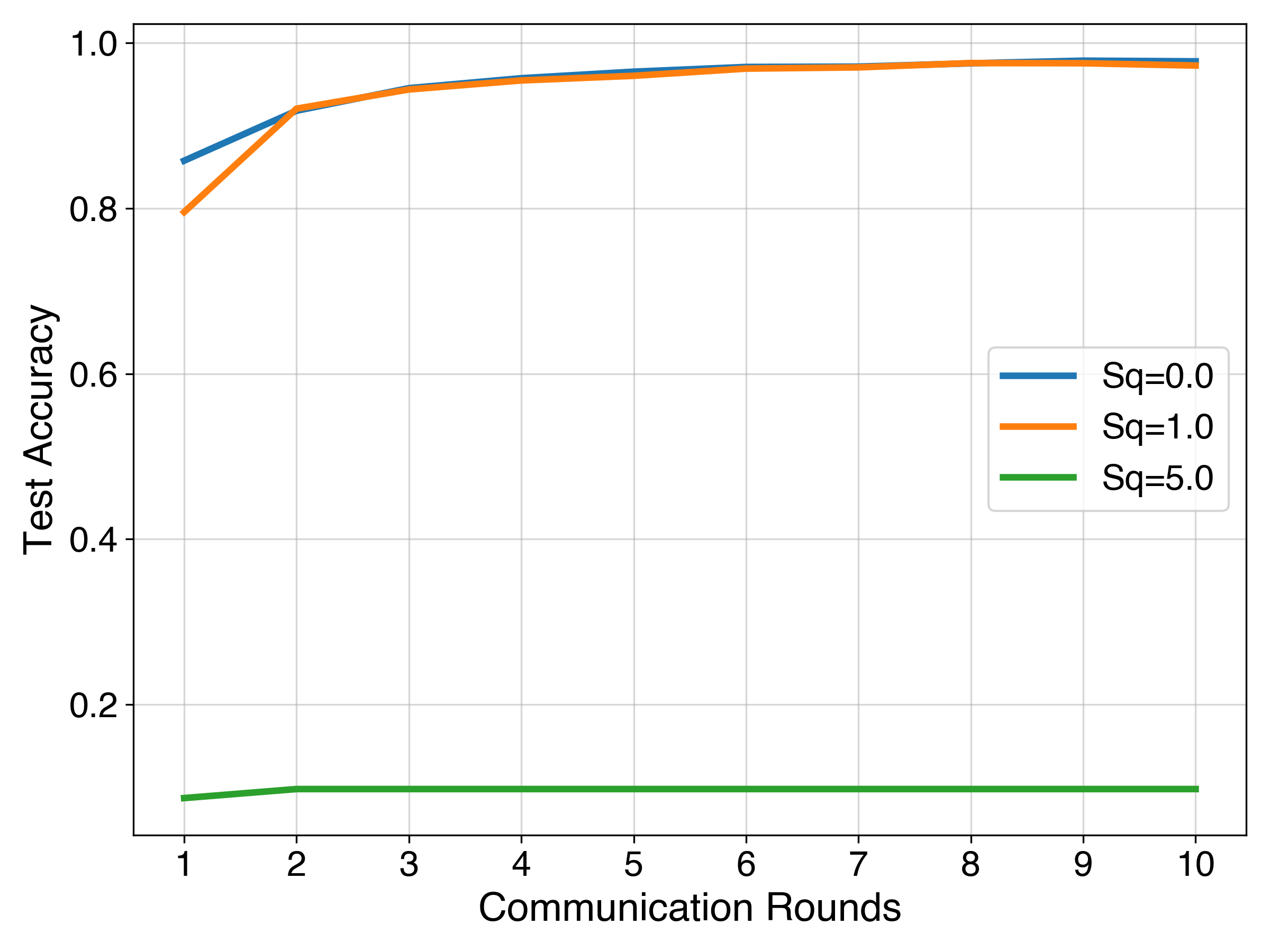}
        \caption{Impact of Beam Squint.}
        \label{fig:exp_b}
    \end{subfigure}
    \hfill
    \begin{subfigure}[b]{0.32\textwidth}
        \centering
        \includegraphics[width=\linewidth]{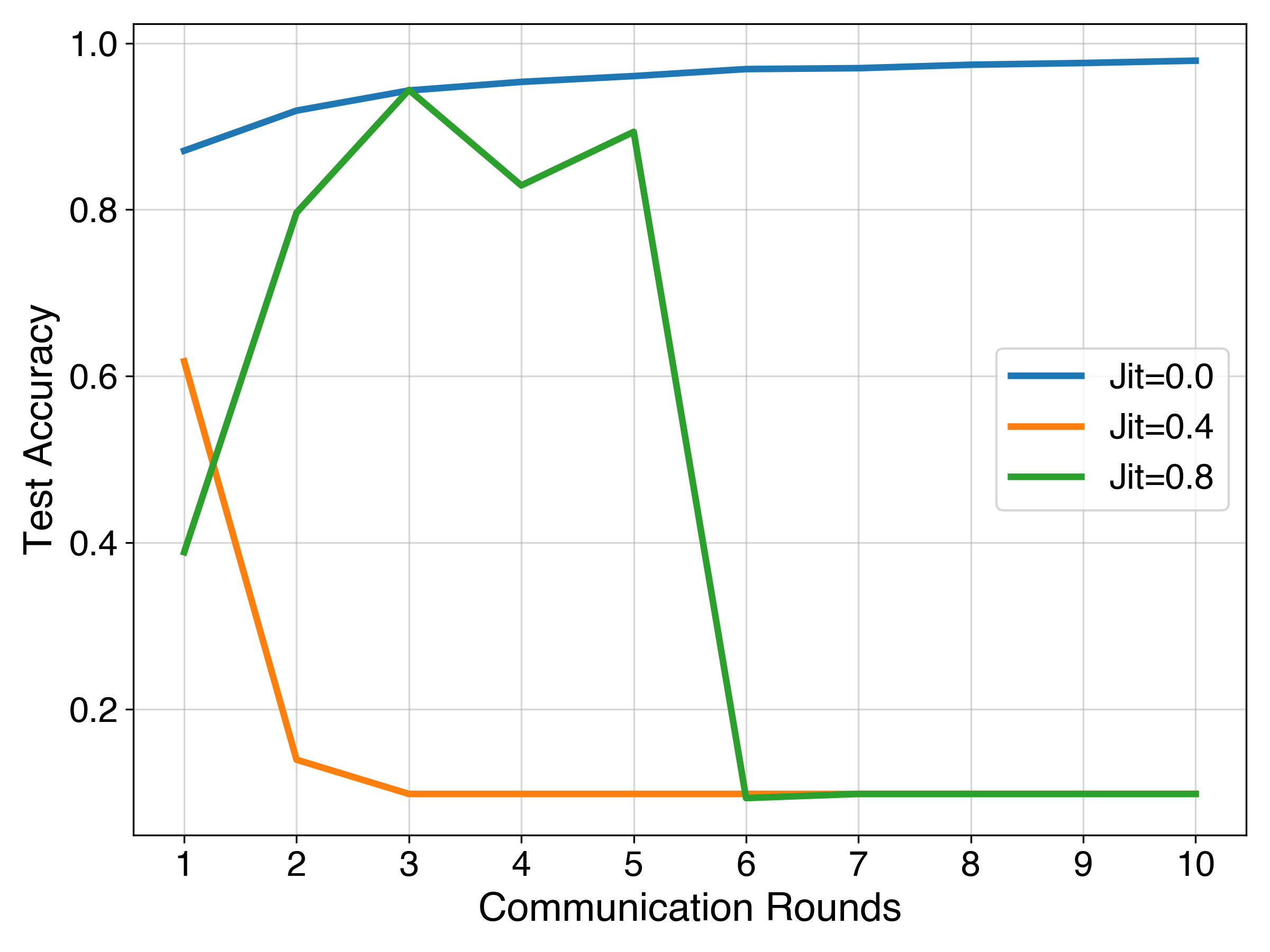}
        \caption{Impact of Jitter.}
        \label{fig:exp_c}
    \end{subfigure}
    \hfill 
    \begin{subfigure}[b]{0.32\textwidth}
        \centering
        \includegraphics[width=\linewidth]{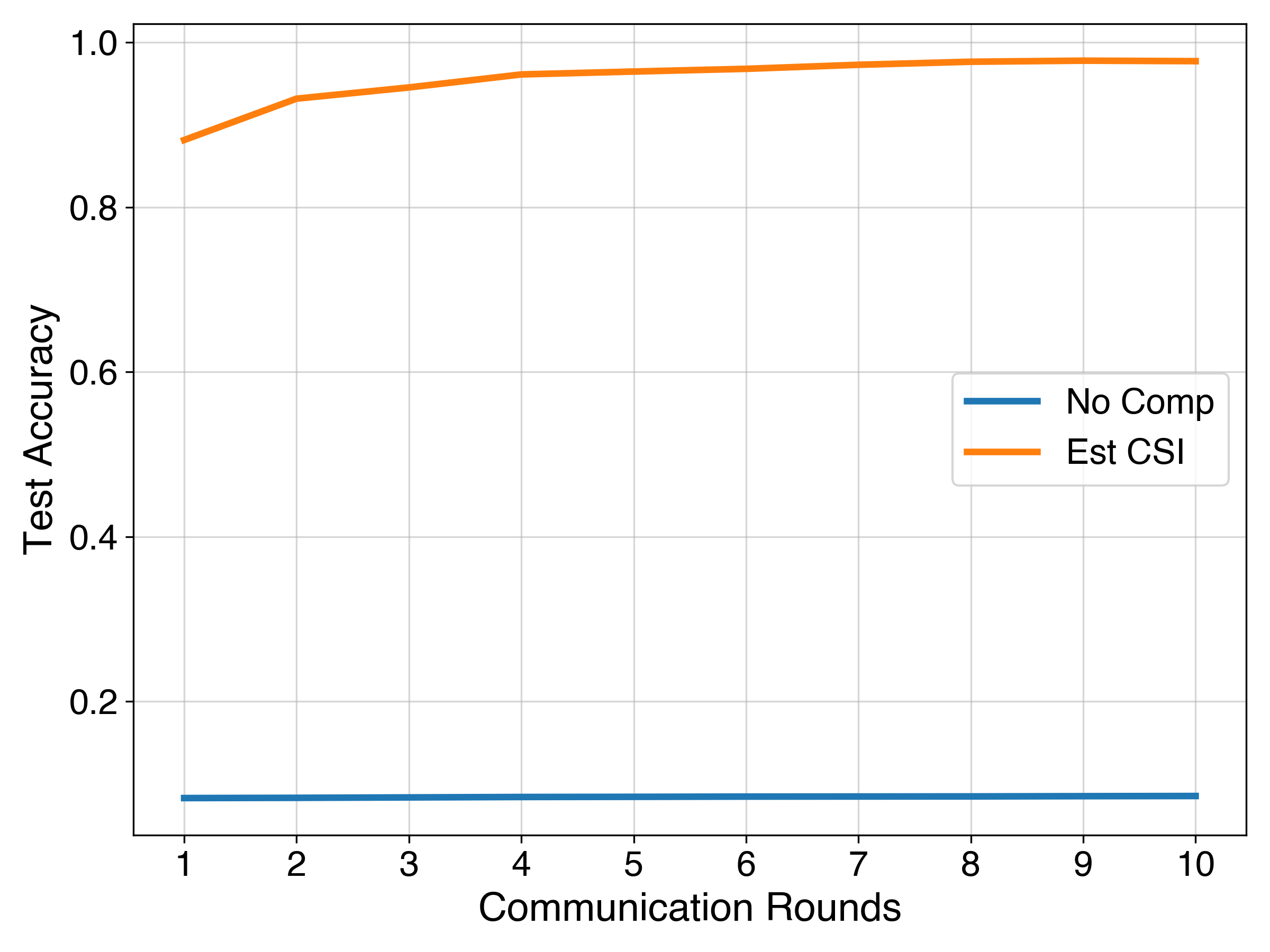}
        \caption{Impact of Compensation.}
        \label{fig:exp_f}
    \end{subfigure}

    \caption{(a) Performance degradation due to beam squint; high severity destroys the link. (b) Jitter stability test showing a cliff effect beyond. (c) Comparison of compensated vs. uncompensated reception, proving equalization is required for convergence.}
    \label{fig:phy_layer_analysis}
\end{figure*}

\begin{figure*}[htbp] 
    \centering
    
    \begin{subfigure}[b]{0.32\textwidth}
        \centering
        \includegraphics[width=\linewidth]{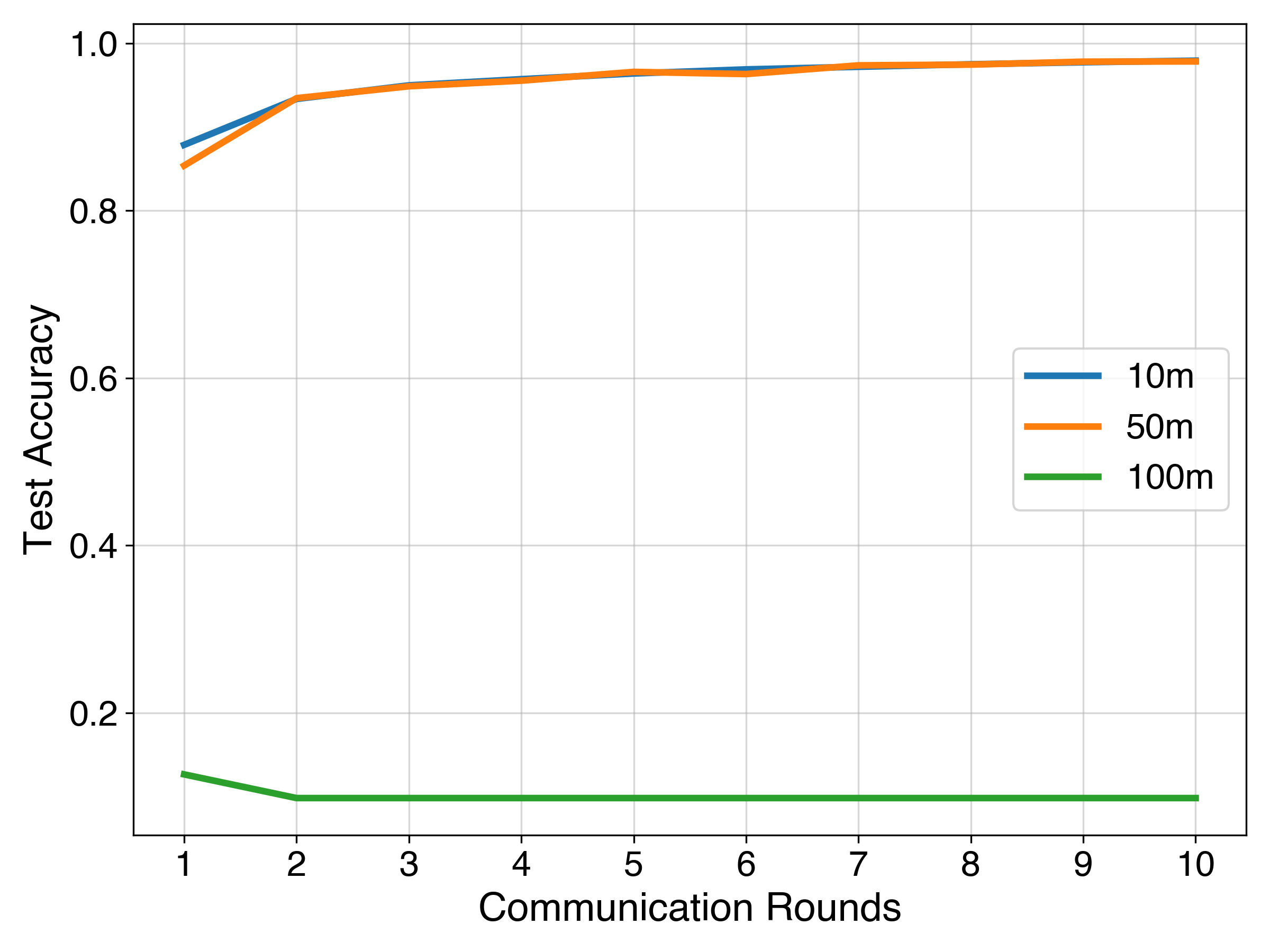}
        \caption{Impact of Distance.}
        \label{fig:exp_h}
    \end{subfigure}
    \hfill 
    \begin{subfigure}[b]{0.32\textwidth}
        \centering
        \includegraphics[width=\linewidth]{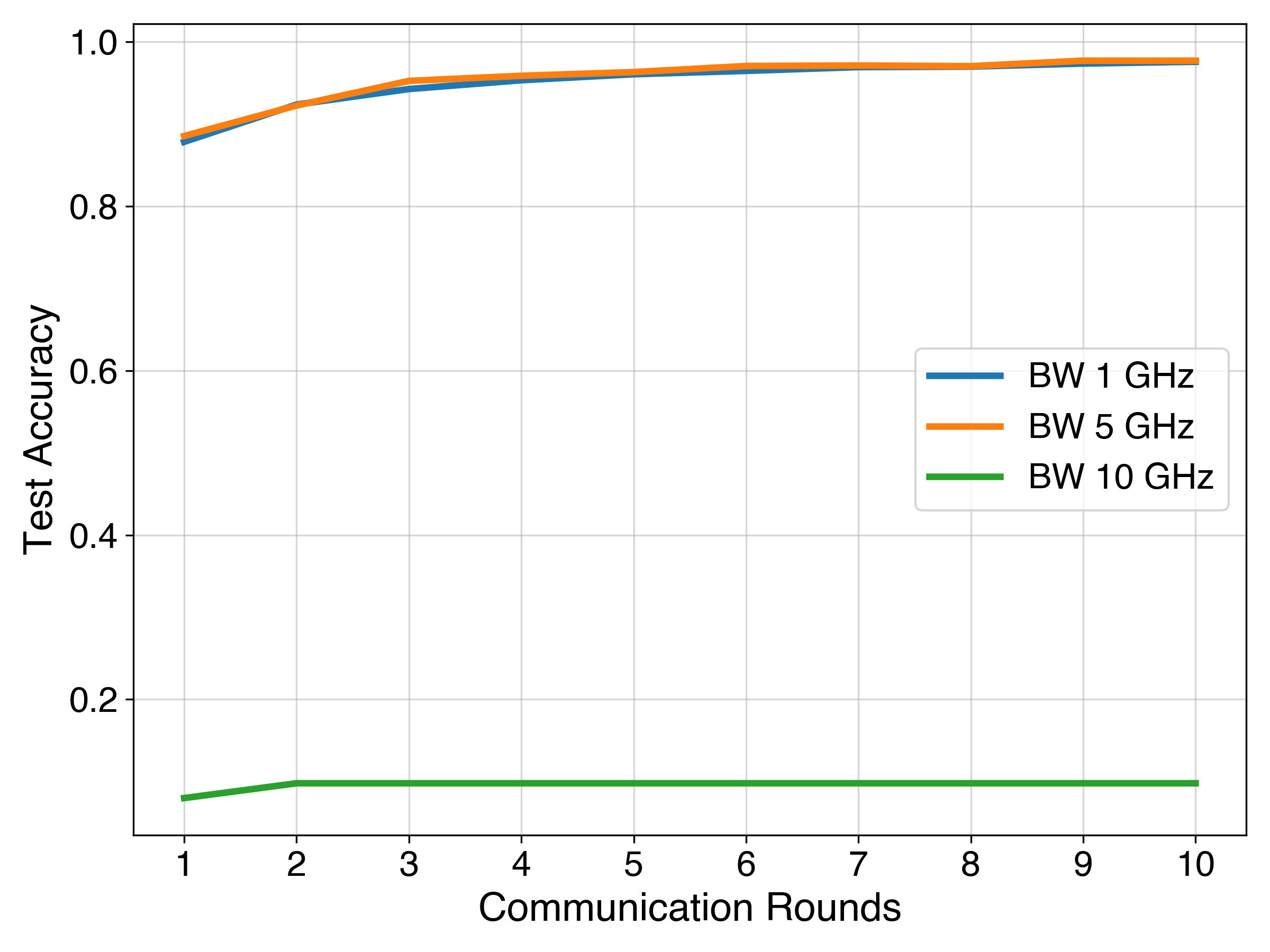}
        \caption{Impact of Bandwidth Expansion.}
        \label{fig:exp_i}
    \end{subfigure}
    \hfill 
    \begin{subfigure}[b]{0.32\textwidth}
        \centering
        \includegraphics[width=\linewidth]{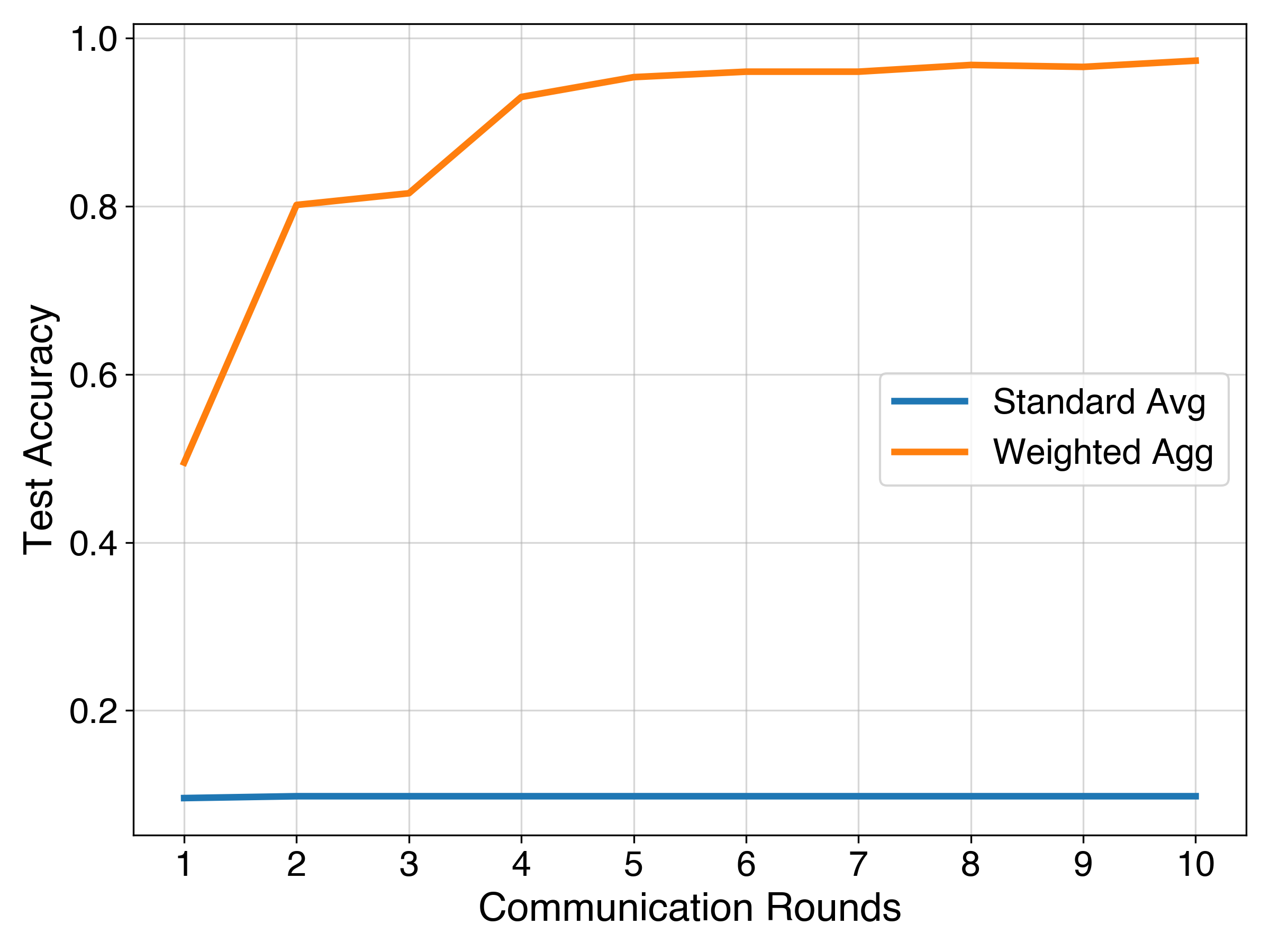}
        \caption{Impact of Weighted Aggregation.}
        \label{fig:exp_k}
    \end{subfigure}
    
    \caption{(a) Client Distance on Convergence; the severe path loss and absorption at high distance degrade the SNR below the erasure threshold, preventing model training. (b) Bandwidth expansion; increasing bandwidth without power scaling raises the noise floor, degrading SNR. (c) Weighted Aggregation vs. FedAvg under severe channel stress.}
    \label{fig:robustness_analysis}
\end{figure*}

\subsection{Beam Squint Severity}
We investigate the frequency-dependent beam split effect by varying the squint severity parameter $s \in \{1.0, 3.0, 5.0\}$, where $s$ acts as a scaling factor for the phase deviation across subcarriers. Fig.~\ref{fig:exp_b} illustrates that the system is robust to moderate squint ($s=1.0$), retaining high accuracy. However, as the severity increases to $s=5.0$, the array gain at the band edges collapses, causing the effective SNR to drop below the safety threshold. The collapse at high squint severity is related to the harmonic mean bottleneck: as more subcarriers are pushed into near-null gains, the harmonic mean of SNRs collapses even though the center-frequency SNR remains high.

\subsection{Jitter and Pointing Stability}
Mechanical misalignment is modeled by introducing random pointing jitter with standard deviation $\sigma_{jit} \in \{0.0, 0.5, 0.8\}$. We observe a ``cliff effect" in Fig.~\ref{fig:exp_c}. The system maintains performance for $\sigma_{jit} \le 0.2$, but accuracy plummets to random guessing at $\sigma_{jit} = 0.4$. This confirms that THz links are extremely sensitive to pointing errors. Robust beam tracking or hybrid beamforming algorithms are essential to maintain alignment within the sub-degree tolerance of high gain antennas.

\subsection{Necessity of Equalization}
We compare the system performance with and without Channel State Information (CSI) compensation. Fig.~\ref{fig:exp_f} demonstrates that uncompensated reception fails to converge ($\approx 14\%$ accuracy). The distortions introduced by the frequency-selective channel scramble the model weights. In contrast, Estimated CSI restores model performance. Blind aggregation is infeasible in THz bands; at minimum, estimated scalar equalization is required at the receiver.

\subsection{Impact of Communication Distance}
We investigate the coverage limits by varying the client-server distance $d \in \{10, 50, 100\}$ meters. The results, illustrated in Fig. \ref{fig:exp_h}, reveal a sharp performance boundary characteristic of high-frequency propagation. At short-to-medium ranges, the system maintains robust convergence, achieving $\approx 98\%$ test accuracy. However, at $d=100$ m, the performance collapses completely to random guessing. This drastic failure is attributed to the combined effect of Friis spreading loss and molecular absorption at 300 GHz. At 100 meters, the received signal strength drops below the decoding sensitivity threshold ($\tau$), causing near total packet erasure. This confirms that THz-enabled FL is fundamentally range-limited.

\subsection{Fundamental Bandwidth Limit}
We analyze the trade-off between bandwidth and accuracy by varying $B \in \{1 \text{GHz}, 5 \text{GHz}, 10 \text{GHz}\}$ while keeping transmit power constant. Consistent with the theory, increasing bandwidth degrades performance (Fig.~\ref{fig:exp_i}). This is due to the thermal noise power $N = k_B T B$ increasing linearly with bandwidth. At 10 GHz, the noise floor rises by 10 dB compared to 1 GHz, dropping the SNR below the decoding threshold. This matches the harmonic mean bottleneck predicted by eq.~\eqref{eq:design-snr-closed}: as bandwidth grows, a handful of edge subcarriers with vanishing gain dominate the sum of inverse SNRs, collapsing the harmonic mean even though the average SNR remains moderate. Ultra-wideband systems must scale transmit power proportionally with bandwidth to maintain spectral efficiency.

\subsection{Robustness of Weighted Aggregation}
Finally, we evaluate the proposed SNR-aware weighted aggregation scheme against standard federated averaging (FedAvg) under the stressed channel conditions (squint $s=15.0$, Power $0.2$ W). Fig.~\ref{fig:exp_k} provides the validation of our framework. Under severe stress, FedAvg fails (accuracy $\approx 9.8\%$) because it incorporates highly noisy updates equally with clean ones and the harmonic mean bottleneck. In contrast, the proposed weighted aggregation achieves near-optimal performance ($\approx 97.6\%$). By assigning aggregation weights inversely proportional to the estimated noise variance and jitter, the server effectively filters out corrupted updates. This validates that the proposed method ensures convergence even in hostile THz environments where traditional methods fail.

\section{Conclusion}
\label{sec:conclusions}

This paper presented a theoretical framework for FL over wideband Terahertz channels, explicitly coupling physical-layer impairments with the convergence bounds of SGD. Our analysis yields three critical design insights:

\begin{itemize}
    \item \textit{The Harmonic Mean Bottleneck:} Under standard unbiased aggregation, the global model error floor is driven by the harmonic mean of per-subcarrier SNRs. Consequently, spectral holes caused by beam squint act as dominant noise sources, degrading the global update.
    \item \textit{The Bandwidth Limit:} As a fundamental physical limit, bandwidth expansion can degrade learning. Without proportional power scaling, widening the spectrum integrates more thermal noise and captures low-gain edge subcarriers, increasing the aggregate gradient estimator variance.
    \item \textit{Bias as a Necessity:} Strictly adhering to unbiased aggregation can be impractical in the THz band. To prevent divergence, the system can employ SNR-weighted aggregation, accepting a controllable objective bias to ensure variance stability.
\end{itemize}

Future research should extend this framework to the downlink, analyzing how beam squint affects the broadcasting of the global model. Additionally, exploring hybrid digital-analog schemes, such as AirComp, could offer a way to bypass subcarrier quantization bottlenecks, provided that non-linear amplifier distortions can be managed. 

\appendices

% ----------------------------------------------------------------------
\section{Proof of Lemma~\ref{lem:theta-concentration}}
\label{app:proof-lemma-theta}

\begin{proof}
The concentration bound applies to the scalar estimator on any single subcarrier $n$. 
Let $\widehat \theta_{i,t}^{(n)}$ be the estimator for the effective mean gain on subcarrier $n$, 
denoted $\theta_i^{\star(n)} = \bar d_i \mu_{H,i}^{(n)}$.

Recall from Section~\ref{sec:thzmodel} that
\[
\widehat \theta_{i,t}^{(n)}
=
\frac{1}{M}\sum_{s=t-M+1}^t z_{i,s}^{(n)},
\qquad
z_{i,s}^{(n)}
=
\frac{\langle y_{i,s}^{(n)},u_{i,s}^{(n)}\rangle}{\|u_{i,s}^{(n)}\|^2}.
\]
Since $y_{i,s}^{(n)} = H_{i,s}^{(n)} u_{i,s}^{(n)} + \epsilon_{i,s}^{(n)}$, 
the statistic simplifies to $z_{i,s}^{(n)} = H_{i,s}^{(n)} + \tilde{\epsilon}_{i,s}^{(n)}$, 
where $\tilde{\epsilon}$ is the effective noise.
Define the martingale difference sequence
\[
X_{i,s}
\;\triangleq\;
z_{i,s}^{(n)} - \mathbb E[z_{i,s}^{(n)}\mid\mathcal F_{s-1}],
\qquad
\mathbb E[X_{i,s}\mid\mathcal F_{s-1}] = 0.
\]
By the sub-Gaussian nature of the additive distortion (Assumption 5), 
the partial sums $S_M = \sum X_{i,s}$ form a martingale with sub-Gaussian increments. 
Applying the standard Azuma–Hoeffding concentration inequality \cite{vershynin2018high} gives, for any $\epsilon > 0$:
\[
\mathbb P\!\left(
\left|\frac{1}{M}S_M\right| > \epsilon
\right)
\;\le\;
2\exp\!\left(-\frac{M\epsilon^2}{2\nu_i^2}\right),
\]
where $\nu_i^2$ is the proxy variance derived from $\sigma_\epsilon^2$ and channel variations.
Substituting $\frac{1}{M}S_M = \widehat\theta_{i,t}^{(n)} - \theta_i^{\star(n)}$, we obtain:
\[
\mathbb P\!\left(
|\widehat\theta_{i,t}^{(n)} - \bar d_i\mu_{H,i}^{(n)}| > \epsilon
\right)
\;\le\;
2\exp\!\left(-\frac{M\epsilon^2}{2\nu_i^2}\right).
\]
Thus, the gain estimate for each subcarrier converges exponentially fast 
to the true mean gain as the pilot history $M$ increases.
\end{proof}

% ----------------------------------------------------------------------
\section{Proof of Lemma~\ref{lem:local-energy-explicit}}
\label{app:proof-lemma-energy}

\begin{proof}
This proof bounds the total squared norm of the update vector, $\mathbb E\|\Delta_{i,t}\|^2$. 
Fix a client $i$ and round $t$. The local iterates satisfy
\vspace{-0.5em}
\[
w^{(i)}_{t,k+1} = w^{(i)}_{t,k} - \eta_{\text{loc}} g^{(i)}_{t,k},
\]
where $g^{(i)}_{t,k} = \nabla f_i(w^{(i)}_{t,k}) + \xi_{t,k}^{(i)}$ with $\mathbb E[\xi_{t,k}^{(i)}]=0$ and $\mathbb E\|\xi_{t,k}^{(i)}\|^2 \le \sigma_{\text{sgd}}^2$.
Unrolling the updates gives $\Delta_{i,t} = -\eta_{\text{loc}}\sum_{k=0}^{K-1} (\nabla f_i(w^{(i)}_{t,k}) + \xi_{t,k}^{(i)})$.

We square the norm and apply the inequality $\|a+b\|^2 \le 2\|a\|^2 + 2\|b\|^2$ to separate the deterministic gradient path from the stochastic noise:
\vspace{-0.5em}
\begin{align}
\mathbb E\|\Delta_{i,t}\|^2 
&\le 2\eta_{\text{loc}}^2 \mathbb E \bigg\| \sum_{k=0}^{K-1} \nabla f_i(w^{(i)}_{t,k}) \bigg\|^2 
+ 2\eta_{\text{loc}}^2 \mathbb E \bigg\| \sum_{k=0}^{K-1} \xi_{t,k}^{(i)} \bigg\|^2.
\label{eq:delta-split}
\end{align}

\noindent Bounding the Noise Term:
Since the mini-batch noise $\xi_{t,k}^{(i)}$ is independent across steps $k$:
\vspace{-0.5em}
\begin{equation}
2\eta_{\text{loc}}^2 \mathbb E \bigg\| \sum_{k=0}^{K-1} \xi_{t,k}^{(i)} \bigg\|^2 
= 2\eta_{\text{loc}}^2 \sum_{k=0}^{K-1} \mathbb E\|\xi_{t,k}^{(i)}\|^2 
\le 2\eta_{\text{loc}}^2 K \sigma_{\text{sgd}}^2.
\end{equation}

\noindent Bounding the Gradient Term:
We use Jensen's inequality $\|\sum_{k=0}^{K-1} x_k\|^2 \le K \sum_{k=0}^{K-1} \|x_k\|^2$ on the first term of \eqref{eq:delta-split}:
\[
2\eta_{\text{loc}}^2 \mathbb E \bigg\| \sum_{k=0}^{K-1} \nabla f_i(w^{(i)}_{t,k}) \bigg\|^2 
\le 
2\eta_{\text{loc}}^2 K \sum_{k=0}^{K-1} \mathbb E \|\nabla f_i(w^{(i)}_{t,k})\|^2.
\]
We approximate the local gradient using the global gradient and heterogeneity. Using $\|a+b\|^2 \le 2\|a\|^2 + 2\|b\|^2$:
\begin{align}
\|\nabla f_i(w^{(i)}_{t,k})\|^2 
&\le 2\|\nabla f_i(w_t)\|^2 + 2L^2\|w^{(i)}_{t,k} - w_t\|^2 \notag \\
&\le 2(2\|\nabla f(w_t)\|^2 + 2G^2) + 2L^2\|w^{(i)}_{t,k} - w_t\|^2.
\end{align}
Substituting this back into the sum:
\[
\text{Grad.} \!\le \!
2\eta_{\text{loc}}^2 K\! \sum_{k=0}^{K-1}\! \left(\! 4\|\nabla f(w_t)\|^2 \!+\! 4G^2 \!+\! 2L^2 \mathbb E \|w^{(i)}_{t,k} \!-\! w_t\|^2 \!\right).
\]
For the drift term $\|w^{(i)}_{t,k} - w_t\|^2$, standard boundedness analysis \cite{stich2018local} shows that for $\eta_{\text{loc}} \le \frac{1}{2L}$, the drift is of order $O(\eta_{\text{loc}}^2 K^2)$. Consequently, the term $2L^2 \|w^{(i)}_{t,k} - w_t\|^2$ is negligible compared to the $O(1)$ terms $4\|\nabla f(w_t)\|^2$ and $4G^2$ when $\eta_{\text{loc}}$ is small. We absorb this higher-order term into the constants, maintaining the dominant coefficients:
\begin{align}
\text{Grad Term} 
&\le 2\eta_{\text{loc}}^2 K^2 (4\|\nabla f(w_t)\|^2 + 4G^2) \notag \\
&= 8\eta_{\text{loc}}^2 K^2 \|\nabla f(w_t)\|^2 + 8\eta_{\text{loc}}^2 K^2 G^2.
\end{align}
Combining the terms yields the result with constants $a_1=8, a_2=2, a_3=8$.
\end{proof}

\section{Proof of Theorem~\ref{thm:main}}
\label{subsec:proof-main}
\begin{proof}
\emph{Step 1: One-step descent:}
Condition on $\mathcal F_t$. By $L$-smoothness of the objective $f$:
\begin{align}
\mathbb{E}[f(w_{t+1})\mid\mathcal F_t]
&\le
f(w_t)
- \eta\,\big\langle\nabla f(w_t),\,
\mathbb E[\widehat{\Delta}_t\mid\mathcal F_t]\big\rangle
\notag\\[-2pt]
&\quad+
\frac{L\eta^2}{2}\,
\mathbb E\!\left[\|\widehat{\Delta}_t\|^2\mid\mathcal F_t\right].
\label{eq:pf-one-step}
\end{align}
The sign of the inner product assumes $\widehat{\Delta}_t$ represents the negative update direction.

\emph{Step 2: Bias removal:}
With per-subcarrier mean-bias compensation (Section~\ref{sec:thzmodel}), the estimator is unbiased up to negligible terms. Thus $\mathbb E[\widehat\Delta_t\mid \mathcal F_t] \approx \frac{1}{m}\sum_{i\in\mathcal S_t}\mathbb E[\Delta_{i,t}\mid \mathcal F_t]$.

\emph{Step 3: Bounding the second moment (Multicarrier):}
We bound the variance of the aggregated update. Using independence across clients:
\vspace{-0.5em}
\begin{align}
\mathbb{E}\big[\|\widehat{\Delta}_t\|^2\mid \mathcal{F}_t\big]
&\le
\frac{1}{m} \left( V_{\mathrm{loc},t} + V_{\mathrm{het},t} + V_{\mathrm{comm},t} \right).
\end{align}
The communication variance $V_{\mathrm{comm},t}$ sums over all subcarriers. 
Using the definition of the update norm with parameter interleaving $\Pi_t$:
\vspace{-0.5em}
\[
\mathbb{E}[\|\Delta_{i,t}^{\pi,(n)}\|^2] = \frac{1}{N_c}\mathbb{E}[\|\Delta_{i,t}\|^2].
\]
Consequently, the multiplicative variance term sums as:
\begin{align}
\sum_{n=1}^{N_c}\! (\sigma_{H,n}^2 \!+\! \omega) \mathbb{E}\|\Delta_{i,t}^{\pi,(n)}\|^2
\!&=\!
\left(\! \frac{1}{N_c} \sum_{n=1}^{N_c} (\sigma_{H,n}^2 \!+\! \omega) \!\right) \mathbb{E}\|\Delta_{i,t}\|^2.
\end{align}
This allows us to factor out the arithmetic mean of the channel parameters. The additive noise term sums separately as $\sum_n \frac{\sigma_{\epsilon,n}^2}{(\mu_{H,i}^{(n)})^2}$.
Inserting Lemma~\ref{lem:local-energy-explicit} (energy bound) into the total variance expression yields the terms in $\Phi_{\mathrm{THz}}^{\mathrm{MC}}$.

\emph{Step 4: Combining descent and variance:}
Substituting the variance bound into eq.~\eqref{eq:pf-one-step} and using the standard local-SGD descent inequality $\langle \nabla f(w_t), \mathbb E[\widehat\Delta_t] \rangle \ge c_0 \eta_{\mathrm{loc}} K \|\nabla f(w_t)\|^2 - c_{\mathrm{drift}} \eta_{\mathrm{loc}}^2 K^2 G^2$:
\begin{align}
\mathbb{E}[f(w_{t+1})]
&\le
\mathbb{E}[f(w_t)]
-\eta\eta_{\mathrm{loc}}K c_0\,\mathbb{E}\|\nabla f(w_t)\|^2
\notag\\[-2pt]
&\quad+
\frac{L\eta^2}{2m}\,\mathbb E[\Psi_{\mathrm{THz}}^{\mathrm{MC}}]
+\text{(SGD \& Het.\ terms)}.
\end{align}

\emph{Step 5: Absorbing multiplicative inflation:}
The inflation component of $\Phi_{\mathrm{THz}}^{\mathrm{MC}}$ scales with $\mathbb E\|\nabla f(w_t)\|^2$. 
Specifically, it is proportional to the mean distortion factor multiplied by the gradient coefficient $a_1$ from Lemma~\ref{lem:local-energy-explicit}:
\vspace{-0.5em}
\[
\frac{1}{m} \left( \frac{1}{N_c} \sum_{n=1}^{N_c} (\sigma_{H,n}^2 + \omega) \right) \cdot a_1 \eta_{\mathrm{loc}}^2 K^2 \mathbb E \|\nabla f(w_t)\|^2.
\]
We assume the stepsize is sufficiently small such that this noise amplification term is $\le \frac{1}{2} c_0 \eta \eta_{\mathrm{loc}} K$, effectively being absorbed into the descent term without destroying convergence.

\emph{Step 6: Telescoping:}
Summing over $t=0 \dots T-1$ and rearranging yields the final rate in eq.~\eqref{eq:nonconvex-rate}.
\end{proof}

\bibliography{references}

@inproceedings{elbir2023federated,
  title={Federated multi-task learning for THz wideband channel and DoA estimation},
  author={Elbir, Ahmet M and Shi, Wei and Mishra, Kumar Vijay and Chatzinotas, Symeon},
  booktitle={2023 IEEE International Conference on Acoustics, Speech, and Signal Processing Workshops (ICASSPW)},
  pages={1--5},
  year={2023},
  organization={IEEE}
}

@ARTICLE{9766110,
  author={Akyildiz, Ian F. and Han, Chong and Hu, Zhifeng and Nie, Shuai and Jornet, Josep Miquel},
  journal={IEEE Transactions on Communications}, 
  title={Terahertz Band Communication: An Old Problem Revisited and Research Directions for the Next Decade}, 
  year={2022},
  volume={70},
  number={6},
  pages={4250-4285},
  doi={10.1109/TCOMM.2022.3171800}}

@ARTICLE{10799091,
  author={Tao, Meixia and Zhou, Yong and Shi, Yuanming and Lu, Jianmin and Cui, Shuguang and Lu, Jianhua and Letaief, Khaled B.},
  journal={Proceedings of the IEEE}, 
  title={Federated Edge Learning for 6G: Foundations, Methodologies, and Applications}, 
  year={2024},
  volume={},
  number={},
  pages={1-39},
  doi={10.1109/JPROC.2024.3509739}}

@article{elbir2022federated,
  title={Federated Learning for THz Channel Estimation},
  author={Elbir, Ahmet M and Shi, Wei and Mishra, Kumar Vijay and Chatzinotas, Symeon},
  journal={arXiv preprint arXiv:2207.06017},
  year={2022}
}

@ARTICLE{7490372,
  author={Han, Chong and Akyildiz, Ian F.},
  journal={IEEE Transactions on Terahertz Science and Technology}, 
  title={Distance-Aware Bandwidth-Adaptive Resource Allocation for Wireless Systems in the Terahertz Band}, 
  year={2016},
  volume={6},
  number={4},
  pages={541-553},
  doi={10.1109/TTHZ.2016.2569460}}

@article{jornet2011channel,
  title={Channel modeling and capacity analysis for electromagnetic wireless nanonetworks in the terahertz band},
  author={Jornet, Josep Miquel and Akyildiz, Ian F},
  journal={IEEE Transactions on Wireless Communications},
  volume={10},
  number={10},
  pages={3211--3221},
  year={2011},
  publisher={IEEE}
}

@article{li2023federated,
  title={Federated learning via over-the-air computation in IRS-assisted UAV communications},
  author={Li, Ruijie and Zhu, Li and Zhang, Guoping and Xu, Hongbo and Chen, Yun},
  journal={Scientific Reports},
  volume={13},
  number={1},
  pages={8009},
  year={2023},
  publisher={Nature Publishing Group UK London}
}

@article{amiri2020federated,
  title={Federated learning over wireless fading channels},
  author={Amiri, Mohammad Mohammadi and G{\"u}nd{\"u}z, Deniz},
  journal={IEEE transactions on wireless communications},
  volume={19},
  number={5},
  pages={3546--3557},
  year={2020},
  publisher={IEEE}
}

@ARTICLE{10726611,
  author={Wang, Jiali and Kaneko, Megumi},
  journal={IEEE Wireless Communications Letters}, 
  title={Exploiting Beam Split-Based Multi-User Diversity in Terahertz MIMO-OFDM Systems}, 
  year={2025},
  volume={14},
  number={1},
  pages={28-32},
  doi={10.1109/LWC.2024.3483808}}

@article{jiang2024terahertz,
  title={Terahertz communications and sensing for 6G and beyond: A comprehensive review},
  author={Jiang, Wei and Zhou, Qiuheng and He, Jiguang and Habibi, Mohammad Asif and Melnyk, Sergiy and El-Absi, Mohammed and Han, Bin and Di Renzo, Marco and Schotten, Hans Dieter and Luo, Fa-Long and others},
  journal={IEEE Communications Surveys \& Tutorials},
  volume={26},
  number={4},
  pages={2326--2381},
  year={2024},
  publisher={IEEE}
}

@article{han2022terahertz,
  title={Terahertz wireless channels: A holistic survey on measurement, modeling, and analysis},
  author={Han, Chong and Wang, Yiqin and Li, Yuanbo and Chen, Yi and Abbasi, Naveed A and Kuerner, Thomas and Molisch, Andreas F},
  journal={IEEE Communications Surveys \& Tutorials},
  volume={24},
  number={3},
  pages={1670--1707},
  year={2022},
  publisher={IEEE}
}

@article{xue2024survey,
  title={A survey of beam management for mmWave and THz communications towards 6G},
  author={Xue, Qing and Ji, Chengwang and Ma, Shaodan and Guo, Jiajia and Xu, Yongjun and Chen, Qianbin and Zhang, Wei},
  journal={IEEE Communications Surveys \& Tutorials},
  volume={26},
  number={3},
  pages={1520--1559},
  year={2024},
  publisher={IEEE}
}

@article{sharma2025terahertz,
  title={Terahertz communication: State-of-the-art and future directions},
  author={Sharma, Sanjeev and Singya, Praveen K and Deka, Kuntal and Adjih, Cedric and Sharma, Mohit},
  journal={IEEE Open Journal of the Communications Society},
  year={2025},
  publisher={IEEE}
}

@article{mahmood2024analysis,
  title={Analysis of terahertz (THz) frequency propagation and link design for federated learning in 6G wireless systems},
  author={Mahmood, Atif and Kiah, Miss Laiha Mat and Azizul, Zati Hakim and Azzuhri, Saaidal Razalli},
  journal={IEEE Access},
  volume={12},
  pages={23782--23797},
  year={2024},
  publisher={IEEE}
}

@book{vershynin2018high,
  title={High-dimensional probability: An introduction with applications in data science},
  author={Vershynin, Roman},
  volume={47},
  year={2018},
  publisher={Cambridge university press}
}

@article{stich2018local,
  title={Local SGD converges fast and communicates little},
  author={Stich, Sebastian U},
  journal={arXiv preprint arXiv:1805.09767},
  year={2018}
}

@article{wang2019beam,
  title={Beam squint and channel estimation for wideband mmWave massive MIMO-OFDM systems},
  author={Wang, Bolei and Jian, Mengnan and Gao, Feifei and Li, Geoffrey Ye and Lin, Hai},
  journal={IEEE transactions on signal processing},
  volume={67},
  number={23},
  pages={5893--5908},
  year={2019},
  publisher={IEEE}
}

@article{alistarh2017qsgd,
  title={QSGD: Communication-efficient SGD via gradient quantization and encoding},
  author={Alistarh, Dan and Grubic, Demjan and Li, Jerry and Tomioka, Ryota and Vojnovic, Milan},
  journal={Advances in neural information processing systems},
  volume={30},
  year={2017}
}

@book{goldsmith2005wireless,
  title={Wireless communications},
  author={Goldsmith, Andrea},
  year={2005},
  publisher={Cambridge university press}
}

@article{akyildiz2014terahertz,
  title={Terahertz band: Next frontier for wireless communications},
  author={Akyildiz, Ian F and Jornet, Josep Miquel and Han, Chong},
  journal={Physical communication},
  volume={12},
  pages={16--32},
  year={2014},
  publisher={Elsevier}
}

@inproceedings{mcmahan2017communication,
  title={Communication-efficient learning of deep networks from decentralized data},
  author={McMahan, Brendan and Moore, Eider and Ramage, Daniel and Hampson, Seth and y Arcas, Blaise Aguera},
  booktitle={Artificial intelligence and statistics},
  pages={1273--1282},
  year={2017},
  organization={PMLR}
}

@ARTICLE{8610080,
  author={Boulogeorgos, Alexandros-Apostolos A. and Papasotiriou, Evangelos N. and Alexiou, Angeliki},
  journal={IEEE Access}, 
  title={Analytical Performance Assessment of THz Wireless Systems}, 
  year={2019},
  volume={7},
  number={},
  pages={11436-11453},
  doi={10.1109/ACCESS.2019.2892198}}

\bibliographystyle{IEEEtran}

\end{document}